\documentclass[11pt]{article}

\usepackage{color}

\usepackage{cite}
\usepackage{authblk}
\usepackage{url}
\usepackage{graphicx}
\usepackage{graphics,epsfig}
 \usepackage{amssymb}
 \usepackage{ulem}
 \usepackage{multirow}
  \usepackage{color}
  \usepackage{subfigure}
  \usepackage{subfig}
 \usepackage{amsmath}
\usepackage[margin=1in]{geometry}
\usepackage[numbers,sort&compress]{natbib}
\usepackage{enumitem}
\usepackage{comment}

\newcommand{\bfl}{\begin{flushleft}}
\newcommand{\efl}{\end{flushleft}}
\newcommand{\bea}{\begin{eqnarray}}
\newcommand{\eea}{\end{eqnarray}}
\newcommand{\be}{\begin{equation}}
\newcommand{\ee}{\end{equation}}
\newcommand{\ben}{\begin{enumerate}[itemsep=0pt,parsep=0pt]}
\newcommand{\een}{\end{enumerate}}
\newcommand{\bi}{\begin{itemize}}
\newcommand{\ei}{\end{itemize}}
\newcommand{\tben}{\begin{enumerate}[itemsep=0.0in,parsep=0.0in]}
\newcommand{\teen}{\end{enumerate}}

\newcommand{\gev}{~\mbox{GeV}}
\newcommand{\mev}{~\mbox{MeV}}

\newcommand{\rhos}{\rho_\sigma}
\newcommand{\rhor}{\rho_r}

\newcommand{\rhom}{\rho_\chi}

\newcommand{\gam}{\Gamma_\sigma}
\newcommand{\sgm}{\sigma}
\newcommand{\ds}{\delta_\sigma}
\newcommand{\dr}{\delta_r}
\newcommand{\dx}{\delta_\chi}
\newcommand{\ts}{\theta_\sigma}
\newcommand{\tr}{\theta_r}
\newcommand{\tx}{\theta_\chi}

\newcommand{\pa}{p_{(\alpha)}}
\newcommand{\ra}{\rho_{(\alpha)}}
\newcommand{\wa}{w_{(\alpha)}}
\newcommand{\ca}{c_{(\alpha)}}

\newcommand{\da}{\delta_{(\alpha)}}
\newcommand{\ta}{\theta_{(\alpha)}}
\newcommand{\tN}{\tilde{N}}

\renewcommand{\em}{\it}

\def\nn{\nonumber}
\def\bec{\begin{center}}
\def\eec{\end{center}}
\def\beq{\begin{eqnarray}}
\def\eeq{\end{eqnarray}}
\def\fr{\frac}

\title{Non-thermal Histories and Implications for Structure Formation}
\author{JiJi Fan}
\author{Ogan \"Ozsoy}
\author[1,2]{Scott Watson}
\affil[1]{Department of Physics, Syracuse University, Syracuse, NY 13244, USA}

\affil[2]{Kavli Institute for Theoretical Physics, Santa Barbara, CA 93106, USA}

\date{\today}
\begin{document}
\maketitle

\begin{abstract}
We examine the evolution of cosmological perturbations in a non-thermal post-inflationary history with a late-time matter domination period prior to BBN. Such a cosmology could arise naturally in the well-motivated moduli scenario in the context of supersymmetry (SUSY) -- in particular in models of Split-SUSY.
Sub-horizon dark matter perturbations grow linearly during the matter dominated phase before reheating and can lead to an enhancement in the growth of substructure on small scales, even in the presence of dark matter annihilations. This suggests that a new scale (the horizon size at reheating) could be important for determining the primordial matter power spectrum.  However, we find that in many non-thermal models free-streaming effects or kinetic decoupling after reheating can completely erase the enhancement leading to small-scale structures. In particular, in the moduli scenario with wino or higgsino dark matter we find that the dark matter particles produced from moduli decays would thermalize with radiation and kinetically decouple below the reheating temperature. Thus, the growth of dark matter perturbations is not sustained, and the predictions for the matter power spectrum are similar to a standard thermal history.  We comment on possible exceptions, but these appear difficult to realize within standard moduli scenarios.  We conclude that although enhanced structure does not provide a new probe for investigating the cosmic dark ages within these models, it does suggest that non-thermal histories offer a robust alternative to a strictly thermal post-inflationary history.
\end{abstract}

\thispagestyle{empty}

\newpage
\tableofcontents
\newpage

\section{Introduction}
Cosmological observations have led to an impressive level of constraint on inflationary model building. However, the post-inflationary universe prior to Big Bang Nucleosynthesis (BBN) remains elusive.  The lack of direct observations at this time is unfortunate, since it is precisely during this epoch that we would hope to probe Beyond the Standard Model (BSM) physics. Even though direct probes on this cosmic period are lacking, we can try and establish some aspects of BSM physics by understanding how new particles and fields may change the expansion history and perhaps alter the inflationary seeds that led to the growth of structure.  

In this paper we investigate non-thermal cosmologies and the effects they can have on the growth of density perturbations during the cosmic dark ages -- the post-inflationary universe prior to BBN. These non-standard cosmologies (in particular those associated with high-scale supersymmetry) are motivated by both fundamental theory \cite{Acharya:2008bk,Acharya:2009zt,Watson:2009hw} and experimentally given rising tensions for natural BSM models~\cite{Kribs:2013lua,Arvanitaki:2013yja,Evans:2013jna, Fan:2014txa, Gherghetta:2014xea}.  Past investigations into the implications of a non-thermal post-inflationary history on cosmological perturbations have already demonstrated there can be important consequences for interpreting Cosmic Microwave Background (CMB) observations and for the restrictions CMB observations place on inflationary model building \cite{Easther:2013nga,Iliesiu:2013rqa}.  Here we examine in detail the evolution of perturbations during the non-thermal period and address the question of whether the extra matter dominated phase predicted by these models can lead to an enhancement in the growth of structure on small scales.  Because sub-Hubble matter perturbations grow linearly during a matter dominated phase (and only logarithmically during a thermal / radiation dominated phase), this suggests a new scale that could prove interesting for the primordial matter power spectrum.  The relevance of this scale for determining the smallest allowed primordial dark matter (DM) structures depends on the reheat temperature at the end of the non-thermal phase, as well as on the free streaming length and kinetic decoupling of the DM.  In this paper we address all of these issues within non-thermal cosmologies and establish in which situations interesting phenomenology may result.  In addition to the general consideration, we also investigate the particular scenario with neutralino DM and heavy moduli in the context of Split Supersymmetry~\cite{Wells:2003tf, ArkaniHamed:2004fb, ArkaniHamed:2004yi, Giudice:2004tc, Arvanitaki:2012ps, Hall:2011jd, Hall:2012zp, ArkaniHamed:2012gw}. 

The paper is organized as follows.  In Section \ref{Bckg} we present a brief review of non-thermal cosmologies and establish the background evolution.  In Section \ref{perts} we present a general discussion of the evolution of cosmological perturbations in these non-thermal cosmologies. We discuss in detail how the extra matter dominated phase can alter both sub-Hubble and super-Hubble matter and radiation perturbations during this time.  We also discuss the different production mechanisms for DM in non-thermal cosmologies and how this relates to expectations for whether structure should be enhanced or suppressed.  One key result from this section is the emergence of a new scale associated with the Hubble radius at the end of the non-thermal phase, which suggests a new possible minimal scale for the smallest allowed primordial DM structures. In Section \ref{decoupling} we compare this scale with other important effects for removing DM structure on small scales, namely the effects of free streaming and kinetic decoupling.  Within our discussion we also discuss how the scalar decay at the end of the non-thermal history can lead to a free-streaming effect that must be taken into account when establishing the relevant scale for the smallest substructures. In Section \ref{split} we consider  neutralino DM in the moduli scenario and discuss whether an enhanced growth of small scale structure is a natural expectation in this scenario.  In Section \ref{conclude} we conclude and relegate more technical details of our analysis to the appendices.

We note that some of our results have overlap with existing papers found in the literature. Many of our results in the perturbation analysis have overlap with that of Erickcek and Sigurdson in \cite{Erickcek:2011us}.  However, we have included the effect of DM annihilations and considered a broader class of non-thermal cosmologies -- as we discuss in Section \ref{Bckg}. We will also emphasize that after reheating, the scattering of DM off radiation could couple DM to radiation and thus wipe out the matter perturbation growth. In summary, we consider the effect of interactions between DM particles and between DM and radiation, which are generally non-negligible in well-motivated particle DM models. We also try to emphasize closely the connection to the microscopic parameters of the underlying theory, which helps to establish which parameter regions prove most relevant. For our considerations of SUSY neutralinos in Section \ref{split} we note the work of Arcadi and Ullio in \cite{Arcadi:2011ev} where they considered strictly wino DM in the context of the $G2$-MSSM.

\section{Non-thermal Cosmologies \label{Bckg}}
In this section we begin by reviewing non-thermal cosmologies and their implications for the primordial DM abundance.
We then present the background equations to model the non-thermal epoch, to be followed in the next section with a study of the perturbations.

There are two assumptions leading to a non-thermal history following inflationary reheating; the existence of shift symmetric scalars (or moduli), and both high and low energy 
sources that break that symmetry.  The former is a generic expectation of BSM physics, whereas low-scale symmetry breaking is motivated by the hierarchy problem and 
inflation provides a gravity mediated source of breaking at the high scale\footnote{In fact, further motivation is provided by inflation itself, where a shift symmetry for the inflaton is 
not enough to obtain adequate inflation, but one must introduce an additional symmetry (such as SUSY) to protect the flatness of the potential against corrections.  In the SUSY case, this leads to an additional scalar
playing the same role as the moduli that we are discussing here (see e.g. \cite{Craig:2014rta}). } \cite{Acharya:2009zt,Watson:2009hw}.  Given these assumptions the scalar will typically be 
displaced from its low energy minimum and its oscillations can lead to an effectively matter dominated universe (see e.g. \cite{Acharya:2008bk} and references within).
For moduli with masses around $100$~TeV and which decay through gravitationally suppressed couplings, this will lead to a late stage of reheating shortly before the time of BBN \cite{Acharya:2008bk}.  Since oscillations begin roughly when $H \sim m \sim 100$ TeV, this implies a long period of matter domination prior to BBN and a modification to the usually assumed radiation dominated post-inflationary universe.  

Depending on the specifics of the non-thermal history (the exact couplings and masses of the fields) there are a few possible predictions for the primordial origin of DM. 
If the energy density of oscillations remains subdominant compared to radiation, this can lead to interesting cosmological predictions \cite{Iliesiu:2013rqa}, but the cosmic evolution will remain thermal.
This will not lead to any change in the growth of structure, so for the remainder of the paper we will assume this is not the case.  
Moreover, top-down approaches to model building typically imply that the moduli will come to dominate the energy density almost immediately following the onset of oscillations \cite{Acharya:2008bk}.
Given that the moduli dominate at the time of decay, this implies a large generation of entropy and so any previous DM abundance will be diluted.  
\\

This leads to the following possible cases \cite{Gelmini:2006pw}:
\vspace{-0.05in}
\bi
\item \underline{Branching Scenario}:  In this case the moduli decay into radiation (standard model particles) and DM particles with no DM annihilations occurring during the process.
The final abundance of DM will then be the (diluted) primordial amount $\sim \Omega_\chi^{(0)} (T_r / T_f)^3$ where $T_r$ and $T_f$ are the reheat and freeze-out temperatures, respectively ($T_r < T_f$), and the decays can lead to a non-thermal source of DM $\Omega_\chi^{NT} \sim B_\chi \rho_\sigma m_\chi / (m_\sigma \rho_c)$ where $B_\chi$ is the branching ratio, $\rho_\sigma \sim H^2 m_p^2$ is the energy density of the moduli at decay, $\rho_c \sim H_0^2 m_p^2$ is the critical density today, and $m_\chi$ and $m_\sigma$ are the masses of the DM and the moduli, respectively.   Within this scenario there is the possibility that the branching ratio could be negligible ($B_\chi \simeq 0$) and so all of the DM is produced during freeze-out before decay\footnote{In this case 
the freeze-out process can actually occur during the matter dominated phase.  This leads to a slightly more involved calculation (e.g. modified freeze-out temperature) than we have presented here 
\cite{Gelmini:2006pw}, however the differences will be irrelevant for our analysis in this paper.}.
Requiring that the non-thermal production provides all of the DM today leads to the constraint \cite{Gelmini:2006pw}
\be
B_\chi=6.4 \times 10^{-8} \left( \frac{5 \; \mbox{MeV}}{T_r} \right) \left( \frac{10.75}{g_{*s}}\right)^{1/3},
\ee
which we see is quite suppressed for low reheat temperatures.  

\item \underline{Annihilation Scenario}:  In this case when the DM is produced from the moduli decay, the abundance results in enough DM so that rapid self annihilations of the DM is possible.  In this case one typically finds that the abundance of DM is primarily of non-thermal origin and the amount of DM today is then $\Omega_\chi^{NT} \sim \Omega_\chi^{std} (T_f / T_r)$.  That is, because of the annihilations the abundance is related to the standard thermal result $\Omega_\chi^{std}$ except that the freeze-out temperature is replaced by the reheating temperature.
Requiring that this provides the totality of DM today forces a relationship between the reheat temperature and DM annihilation cross-section (see e.g. \cite{Fan:2013faa}).  
For a reheat temperature around $5$ MeV this results in an enhanced DM interaction rate $\langle \sigma v \rangle \sim 10^{-23} \; \frac{\mbox{cm}^3}{\mbox{s}}$ implying the possibility of interesting predictions for the indirect detection of DM~\cite{Cohen:2013ama, Fan:2013faa, Hryczuk:2014hpa}.
\ei

Given these two possible scenarios we next consider the evolution of the cosmological background.  We note that in \cite{Erickcek:2011us} the authors only considered the `branching scenario' where DM annihilations are negligible. Here we extend their analysis to consider both cases, noting that motivation from fundamental theory so far seems to favor the `annihilation scenario'.

\subsection{Background Evolution}
 The treatment of the background equations has appeared in many places in the past, and we find our results to be in close agreement with those of \cite{Giudice:2000ex}.
We are interested in the background evolution following the end of inflationary reheating, assuming
a high-scale model of inflation with reheating temperatures near the GUT scale. 
Once the expansion rate becomes comparable to the moduli mass,
coherent oscillations of the scalar will lead to a matter dominated phase.  Within this regime we can describe 
the cosmological background as a system of three interacting fluids as
\bea
 \dot{\rho}_\sgm &=& -3H\rhos - \gam\rhos, \label{BGE1} \\
\dot{\rho}_ r &=& -4H\rhor +(1-B_\chi) \gam \rhos +\frac{\langle \sigma v \rangle}{m_\chi}\left[\rhom^2-\rho_{\chi,eq}^2\right], \label{BGE2}\\
\dot{\rho}_\chi &=&-3H\rhom +B_\chi \gam\rhos -\frac{\langle \sgm v \rangle}{m_\chi}\left[\rhom^2-\rho_{\chi,eq}^2\right], \label{BGE3}
\eea
where $\gam \sim (m_{\sigma}^3/m_p^2)$ is the decay rate of the scalar with $m_p=2.44 \times 10^{18} \gev$ the reduced Planck mass, $\langle \nn\sgm v \rangle$ is the self annihilation cross section of DM particles with mass $m_\chi$ and $B_\chi$ is the branching ratio for scalar to decay to DM. We assume all other decays result in relativistic particles. We will be interested in the non-relativistic regime of DM $T \ll m_\chi$ and so can neglect the equilibrium terms $\rho_{\chi,eq}^2\sim e^{-m_\chi/T}$ in \eqref{BGE2} and \eqref{BGE3} . The temperature is related to the radiation energy density as $\rhor=  {\pi^2 g_*T^4}/{30}$, and we take care to track the non-standard relation between the temperature and expansion rate during the entropy production within the matter (moduli) dominated phase \cite{Giudice:2000ex}.

The Hubble and Friedmann equations are
\bea
3 H^2 m_p^2 &=& \sum_\alpha \rho_{(\alpha)}, \label{FRW0}\\
2\dot{H} m_p^2&=& -\sum_\alpha ( \rho_{(\alpha)}+p_{(\alpha)}) ,\label{FRW}
\eea
where $\alpha$ runs over the values $\alpha=\sigma,r,\chi$ for each fluid and dot denotes differentiation with respect to cosmological time $t$.  Instead of working with time it is convenient to express the equations in the number of e-folds, $H dt= dN=d(\ln{a})$, so that the dynamical equations \eqref{BGE1}-\eqref{BGE3} and (\ref{FRW}) become
\bea\label{BEGN}
\label{rhos}\fr{d\rhos}{dN}  &=& -3\rhos - \fr{\gam}{H}\rhos, \\
\label{rhm}\fr{d\rhom}{dN} &=&-3\rhom +B_\chi \fr{\gam}{H} \rhos -\frac{\langle \sgm v \rangle}{m_\chi H}\rhom^2 , \\
\label{rhr}\fr{d\rhor}{dN} &=& -4\rhor +(1-B_\chi) \fr{\gam}{H} \rhos +\frac{\langle \sigma v \rangle}{m_\chi H}\rhom^2,\\
\label{H}\fr{dH}{dN}&=&-\fr{1}{2Hm_p^2}(\rhos+\rhom+\fr{4}{3}\rhor),
\eea
subject to the energy constraint \eqref{FRW0}.
 
We begin studying the behavior of the system well within the matter dominated phase resulting from the coherent oscillations of the moduli, i.e. $t  \sim H^{-1} \gg  m_\sigma^{-1}$.  Moduli decays into both DM (which is by this time non-relativistic) and radiation do not significantly reduce the abundance of moduli until the time of decay $ t_d \sim H^{-1} \simeq \gam^{-1} $, however the decays do affect the scaling behavior as discussed in e.g. \cite{Giudice:2000ex}.
Indeed, we find that prior to reheating the moduli evolve as expected but that the DM and radiation scale differently 
\bea
\label{rhss}\rhos(N) &\simeq & \rho_{\sgm}^{(0)}~e^{-3N},\\
\label{rhms}\rhom(N) &=& \rho_{\chi}^{(0)}~e^{-3N/2},\\
\label{rhrs}\rhor(N) &=& \rho_{r}^{(0)}~e^{-3N/2},
\eea  
where we choose initial values so that $ \rho_{\chi}^{(0)}, \rho_{r}^{(0)}<<\rho_{\sgm}^{(0)}$ and DM will be primarily of non-thermal origin\footnote{In explicitly constructed models with moduli and TeV scale (gravity or anomaly mediated) SUSY breaking this is typically found to be the case -- see e.g. \cite{Acharya:2008bk}.}. 
The scaling behavior in \eqref{rhms} and \eqref{rhrs} is similar to the case studied recently in  \cite{Erickcek:2011us}, where 
the annihilations of DM were not taken into account. However, (as we have checked numerically) the behavior here is due to a near cancelation between the decay and annihilation terms on the right hand side of equations \eqref{rhm} and \eqref{rhr}, which allows the DM density to track quasi-static equilibrium \cite{Arcadi:2011ev,Cheung:2010gj} and so it dilutes more slowly than the standard $\sim 1/a^3$ as seen in (\ref{rhms}). 
This characterizes the behavior of the system until near $H^{-1} \sim \gam^{-1}$ when the decays become significant enough to reduce the scalar abundance.
The evolution during this time is described well by the sudden decay approximation, and given a large enough yield of DM rapid annihilations will occur -- see \cite{Watson:2009hw} for more details.  

\begin{figure}[t!]
\begin{center}
\includegraphics[scale=0.61]{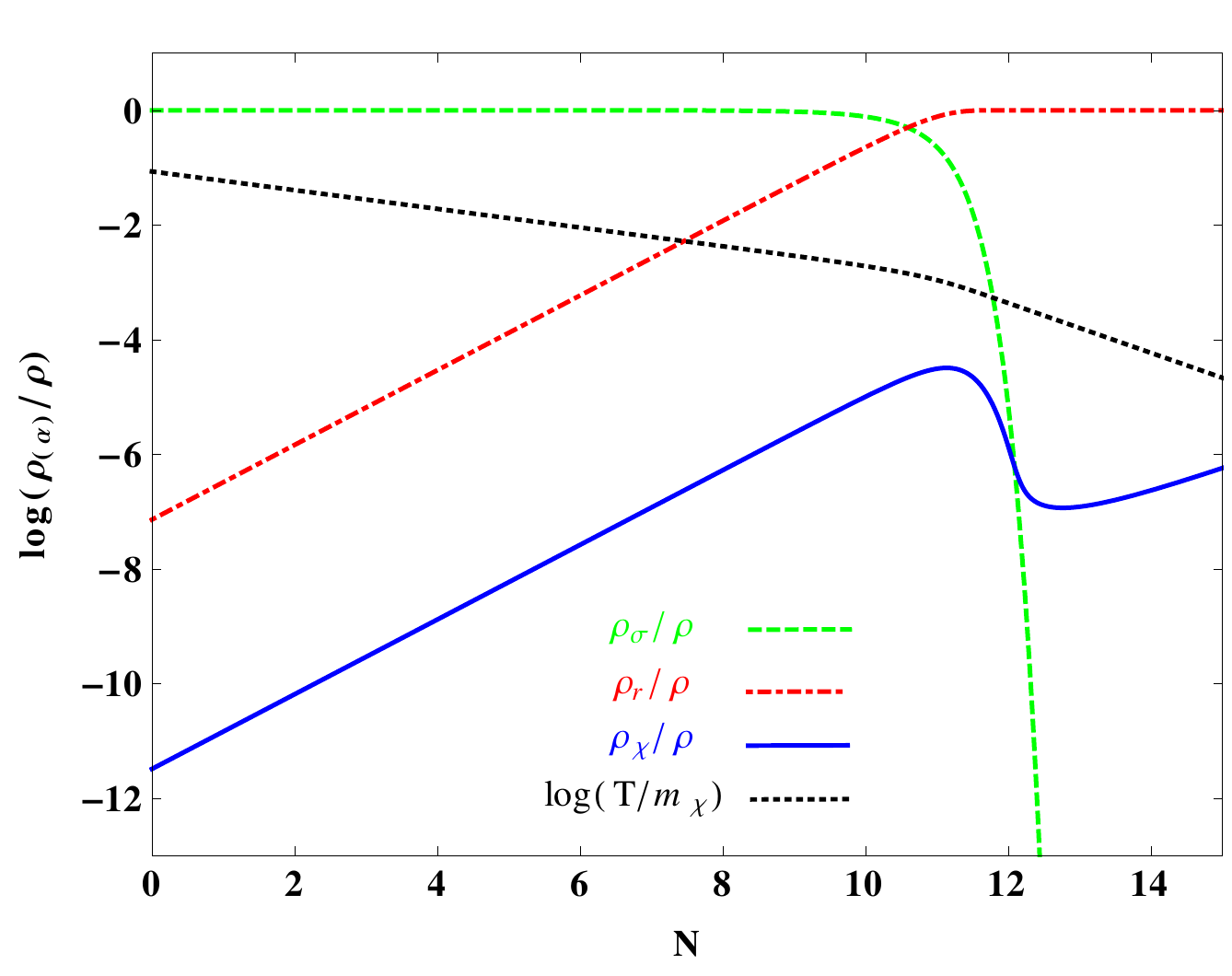}\includegraphics[scale=0.595]{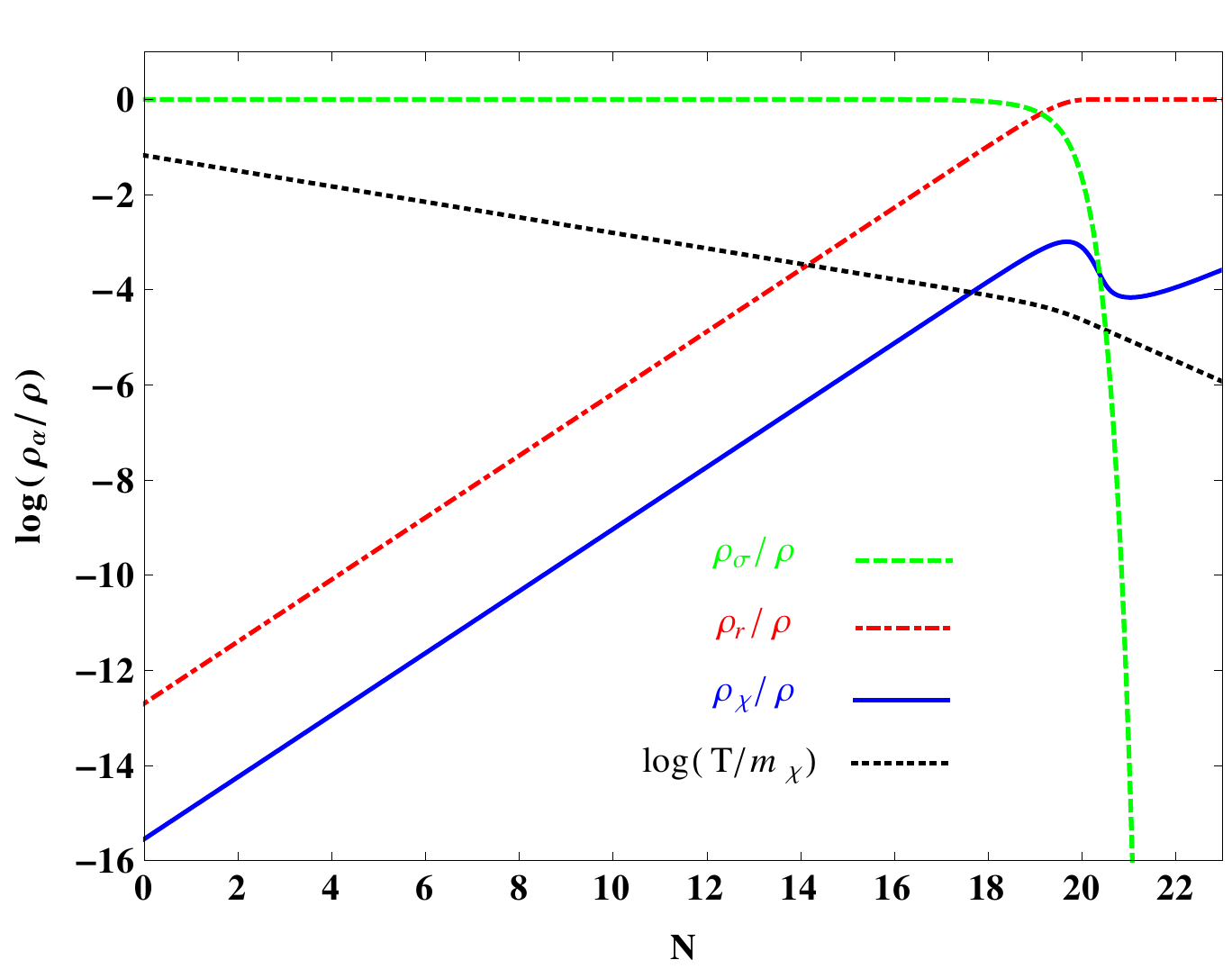}
\end{center}
\caption{Evolution of the background energy densities compared to the total density (as discussed in the text) for two different non-thermal cosmologies.
In both figures we take $m_\chi=500 \gev, B_\chi=1/3$,  $g_*=30$, and $\langle \sigma v \rangle=3\times 10^{-8} \gev^{-2}$. On the left we chose the initial dimensionless decay rate as $\gam/H_0 \simeq 2\times 10^{-7}$ and the moduli mass $m_\sigma=10^6 \gev$. On the left the universe becomes radiation dominated at $N_{rh}\simeq 10.6$ and the reheat temperature is $T_{r}\simeq 707 \mev$. 
Whereas on the right we have  $\gam/H_0 = 0.5\times 10^{-12}$ and $m_\sigma=10^5 \gev$, with $N_{rh}\simeq 19$ at reheating and
$T_{r} \simeq 22 \mev$. \label{fig:BG}}
\end{figure}

The dynamics of the entire system is easily solved numerically, and 
the evolution of the background energy densities as a fraction of total $\rho=\rhos+\rhor+\rhom$ for two different non-thermal cosmologies is presented in 
Figure \ref{fig:BG}.
For both sets of parameters the DM and radiation is found to evolve as $\sim  e^{-3N/2} \sim a^{-3/2}$, until the time of reheating at $H^{-1} \sim \gam^{-1}$.
Then, the scalar energy density becomes exponentially suppressed, $\rhos\sim e^{-2 \gam /3 H(N)}$ and most of the energy density deposited in the coherent scalar oscillations will be transferred to radiation and DM fluids in a very short time interval as seen in both figures above.
The sudden decay will increase the DM density to a critical value such that DM annihilations terms in \eqref{rhm} will be more important than the Hubble expansion terms, resulting in rapid annihilations of DM into radiation until these two terms balance each other. On the other hand, DM pair annihilations do not have an observable effect on the radiation fluid due to the large hierarchy between the energy densities of these fluids at reheating. Once all the energy in scalar oscillations is transferred into DM and radiation, all the source terms in background equations are negligible and the fluids evolve as
$\rhor\sim e^{-4N}$ and $\rhom\sim e^{-3N}$. 
Given both an analytic and numeric description of the system we now turn to a study of the evolution of cosmological perturbations.  

%------------------------------------------------------------%

\section{Cosmological Perturbations \label{perts}}

The evolution equations for the scalar perturbations can be derived by perturbing covariant versions of the background equations presented in Section \ref{Bckg} -- details appear in Appendix A. Consistent with our analysis in that section, we will drop terms the equilibrium terms $\rho_{\chi,eq}$ in equations \eqref{ds}-\eqref{tr} focusing on the evolution after DM has become non-relativistic.  We work in longitudinal gauge where the scalar metric perturbations are
\be
ds^2=-\left(1+2 \Phi \right) dt^2 + a(t)^2\left(1-2 \Psi \right) \delta_{ij} dx^i dx^j.
\ee
In the absence of anisotropic stress for the fluid sources we have $\Phi=\Psi$ and working in momentum space (and suppressing the wave number) the time-time component of the perturbed Einstein equation is
\be
\label{EE1}
\left( \frac{k^2}{3 a^2 H^2} + 1 \right) \Phi + \Phi' =-\fr{1}{6H^2m_p^{2}}\sum_{\alpha} \delta \rho_{(\alpha)}, 
\ee
where prime denotes derivatives with respect to number of e-folds and $v_{(\alpha)}, \delta \rho_{(\alpha)}$, $ \delta p_{(\alpha)}$ are scalar velocity, density and pressure perturbations for each fluid, respectively. Introducing fractional density perturbations $\delta_{(\alpha)} \equiv \delta \rho_{(\alpha)} / \rho_{(\alpha)}$ and defining the velocity perturbation for each fluid as $\theta_{(\alpha)}=a^{-1}\nabla^{2}v_{(\alpha)}$,  the  continuity equations in momentum space are given by
 \bea
\label{ds} \delta'_\sigma+ \fr{\ts}{aH}-3\Phi' &=&-\fr{\gam}{H}\Phi, \\
\label{dx}\delta'_\chi+\fr{\tx}{aH}-3\Phi' &=& B_\chi\fr{\gam}{H}\left( \frac{\rhos}{\rhom} \right)\left[ \ds -\dx+ \Phi \right]-\fr{\langle \sigma v \rangle}{m_\chi H}\rhom\left[\dx + \Phi\right], \\
\label{dr} \delta'_r+\frac{4}{3}\fr{\tr}{aH}-4\Phi'&=&(1-B_\chi)\fr{\gam}{H} \left( \fr{\rhos}{\rhor} \right)\left[\ds -\dr + \Phi \right]+\fr{\langle \sigma v \rangle}{m_\chi H}\left(\fr{\rhom}{\rhor}\right)\rhom\left[2\dx -\dr+ \Phi\right],
\eea
Similarly, the equations for velocity perturbations are
 \bea
\label{ts} \ts' +\ts-\fr{k^2}{aH}\Phi &=& 0, \\
\label{tx} \tx' +\tx -\fr{k^2}{aH}\Phi &=& B_\chi\fr{\gam}{H} \left( \fr{\rhos}{\rhom}\right)\left[ \ts -\tx\right], \\
\tr' -\fr{k^2}{aH}\left(\fr{\dr}{4} +\Phi\right) &=&(1- B_\chi) \fr{\gam}{H} \left( \fr{\rhos}{\rhor}\right) \left[ \frac{3}{4} \ts -\tr\right]+ \fr{\langle \sigma v \rangle}{m_\chi H}\left(\fr{\rhom}{ \rhor}\right)\rhom \label{tr}\left[\fr{3}{4}\tx-\tr\right].
 \eea
We have assumed each fluid has a definite equation of state with $p_{(\alpha)}=w_{(\alpha)}\rho_{(\alpha)}$ in deriving \eqref{ds}-\eqref{tr} and hence $\delta p_{(\alpha)}= c_{s(\alpha)}^{2} \delta\rho_{(\alpha)}$ with $c_{s(\chi)}^2 = c_{s(\sgm)}^2 = 0, c_{s(r)}^2 =1/3$. This set of differential equations can be closed by the perturbed Einstein equation \eqref{EE1}. 

\subsection{Initial Conditions} 
In order to calculate the evolution of perturbations we need to specify initial conditions.  We set these initial conditions are well after the scalar dominated era has begun and when all modes of interest are super-Hubble, $k/aH\to 0$. Given the multiple fluid setup and the presence of decays, one concern may be a substantial contribution to an isocurvature component that could then be in conflict with CMB observations \cite{Lemoine:2009is}.
However, here we will be interested in the case when the modulus completely dominates the energy density before decay, and so any existing isocurvature carried by the moduli will be eliminated as the moduli evolve to dominate -- see \cite{Iliesiu:2013rqa} and references within. Moreover, any DM or radiation that exists prior to moduli domination is found to be insignificant compared to that coming from decay, and so this does not lead to a constraint from observations\footnote{In fact,  it was found in \cite{Iliesiu:2013rqa} that the interesting case corresponds to when the modulus does not completely dominate, and even then it was demonstrated that the importance of isocurvature constraints depends sensitively on the theoretical priors of the model.}.  We elaborate on the role of isocurvature in Appendix B, but given these considerations we are interested in strictly adiabatic initial conditions for the multi-fluid perturbations so that
\be\label{aini}
\fr{\delta\rho^{(0)}_{\alpha}}{\rho'_{\alpha}}=\fr{\delta\rho^{(0)}_{\beta}}{\rho'_{\beta}}.
\ee 

Using the background fluid equations \eqref{rhos}-\eqref{rhr} with the ansatz \eqref{rhss}-\eqref{rhrs} and remembering that during scalar domination $\gam/H\ll 1$, from \eqref{aini} we obtained the following relation for fluid perturbations
\be\label{aini2}
\ds^{(0)}=2\dx^{(0)}=2\dr^{(0)}.
\ee
This relation differs from the standard relation ($\dx^{(0)}=(3/4)\dr^{(0)}$) due to the presence of decays and entropy production. Taking the super-horizon limit $k/aH\to 0$ of \eqref{EE1} in a scalar dominated universe 
$\rhos\gg\rhor,\rhom$, we have 
\be\label{EE1SH}
 \Phi\simeq-\fr{1}{6H^2m_p^{2}}\rhos\ds ,
\ee
where we used that the gravitational potential $\Phi$ is conserved on super-Hubble scales. Since $\rhos\simeq 3H^2m_p^2$ during scalar domination \eqref{EE1SH} implies the following initial condition for long wavelength gravitational perturbations $\ds^{(0)}=-2\Phi_0$ and it follows from \eqref{aini} that 
$\dx^{(0)}=\dr^{(0)}=-\Phi_0$.  Finally, because scalar velocity perturbations quickly decay outside of the Hubble radius, we will set their initial value to vanish
on large scales when solving \eqref{ts}-\eqref{tr}. 

\subsection{Evolution of the Perturbations during Moduli Domination} 
In this section, we examine the evolution of the perturbations for modes that enter the Hubble radius during moduli domination. We note that these modes will be small compared to the size of the horizon at reheating, $k^{-1}< k_{rh}^{-1}$, and thus it will be important for determining the growth of structure at that time.  Our results for this part of the analysis are in general agreement with \cite{Erickcek:2011us}, but here we will include the effect of annihilation terms on the evolution of the perturbations.

\subsubsection{Moduli Perturbations}
We have seen that moduli domination leads to an effectively matter dominated universe, and so the gravitational potential $\Phi$ will be constant on both super and sub-Hubble scales (neglecting the second decaying mode) \cite{Brandenberger:1992dw}. Therefore, we can set $\Phi=\Phi_0$ during the scalar dominated era for both super and sub-Hubble scales.  Using this, we can rewrite \eqref{ts} as 
\be\label{ts*}
\ts' + \ts = \fr{k^2}{H_0}\Phi_0 e^{N/2},
\ee
where we used $H=H_0~e^{-3N/2}$ in a matter dominated Universe.   
This equation can be solved to give the behavior for all wavelengths, concentrating on the growing mode we have    
\be\label{tss}
\ts(k,N)=\fr{2}{3}\fr{k^2}{H_0} \Phi_0 e^{N/2},
\ee
which confirms that long wavelength vector modes are unimportant.
From \eqref{tss}, we can derive the evolution of the scalar perturbation $\ds$ during the  scalar dominated era subject to the initial condition $\ds^{(0)}=-2\Phi_0$. Noting that until the time of reheating we have $\gam/H \ll 1$ and $\Phi$ is constant, we can rewrite \eqref{ds} as
\be\label{dsrw}
\delta'_\sigma(k,N)\simeq - \fr{\ts}{H_0}e^{N/2}
\ee
and using the result in \eqref{tss} we integrate to find
\be\label{dss}
%\ds(k,N) =-2\Phi_0  - \fr{2}{3}\fr{k^2}{H_0^2} \Phi_0 \left(  e^{N} \right) - \fr{2}{3}\fr{\Gamma_\sigma}{H_0^2} \Phi_0 \left(  e^{3N/2}  \right) , 
\ds(k,N)\simeq -2\Phi_0 - \fr{2}{3}\fr{k^2}{H_0^2} \Phi_0 e^{N},
\ee
which is again valid on all scales.

\subsubsection{Dark Matter and Radiation Perturbations}
In the absence of the terms on the right hand side of the fluid perturbation equations \eqref{dx}-\eqref{tr}  the perturbations would just evolve as expected in a matter dominated universe. However, these additional terms will be important during the period of scalar domination and solutions for the complete system can be found by noting that the 
background dependent quantities on the right hand side of these equations are time independent constants.  This can be seen by using the background solutions \eqref{rhss} - \eqref{rhrs} to determine the coefficients on the right hand side of \eqref{dx} which scale as
\bea \label{pref}
 \label{a1} B_\chi\fr{\gam}{H}\left(\frac{\rhos}{\rhom} \right) &\longrightarrow& B_\chi\fr{\gam}{H_0}\left(\frac{\rhos^{(0)}}{\rhom^{(0)}} \right) \equiv A_1, \\
 \label{a2} \fr{\langle \sigma v \rangle}{m_\chi H}\rho_{\chi} &\longrightarrow&  \fr{\langle \sigma v \rangle}{m_\chi H_0}\rho_{\chi}^{(0)} \equiv A_2.
\eea
where have used that $H = H_0~e^{-3N/2}$ during the scalar dominated epoch and $A_1$ and $A_2$ are constants.
While for density perturbations of radiation, from \eqref{dr} and \eqref{tr} we again find that the scaling cancels and the coefficients are determined by their initial values,
\bea \label{pref2}
 \label{a3} (1-B_\chi)\fr{\gam}{H}\left(\frac{\rhos}{\rhor} \right) &\longrightarrow& (1-B_\chi)\fr{\gam}{H_0}\left(\frac{\rhos^{(0)}}{\rhor^{(0)}} \right) \equiv A_3, \\
 \label{a4} \fr{\langle \sigma v \rangle}{m_\chi H} \left( \frac{\rho_{\chi}}{\rhor} \right){\rho_{\chi}} &\longrightarrow&  \fr{\langle \sigma v \rangle}{m_\chi H_0} \left( \frac{\rho^{(0)}_{\chi}}{\rhor^{(0)}} \right){\rho^{(0)}_{\chi}} \equiv A_4.
\eea
Using this information and selecting a range of initial values motivated from SUSY model building we find that the annihilation and decay terms are of comparable importance.
We can first solve for the DM perturbations.
The velocity perturbations of the DM fluid during scalar domination can be found by using \eqref{tss} in \eqref{tx},
\be\label{txe}
\tx' +(1 + A_1)\tx = (1 +\fr{2A_1}{3}) \fr{k^2}{H_0} \Phi_0 e^{N/2}.
\ee
Integrating \eqref{txe}, we find 
\be\label{txs}
\tx(k,N)=\fr{2}{3}\fr{k^2}{H_0} \Phi_0 e^{N/2}.
\ee
Similarly, using \eqref{dss} and \eqref{txs} in \eqref{dx} and remembering that the background coefficients are constants we have
\be
\dx' + (A_1+A_2)\dx =-(A_1+A_2)\Phi_0 - \fr{2}{3}(1+A_1)\fr{k^2}{H_0^2} \Phi_0 e^{N}.
%\dx' + 2A_1 \dx \simeq -2A_1\Phi_0 - \fr{2A_1}{3}\fr{k^2}{H_0^2} \Phi_0 e^{N},
\ee
Integrating the above equation gives
\be\label{dxs}
\dx(k,N)=-\Phi_0 - \fr{2}{3} \left( \frac{1+A_1}{1+A_1+A_2} \right) \fr{k^2}{H_0^2} \Phi_0 e^{N},
%\dx(k,N)\simeq-(\Phi_0 + \fr{1}{3}\fr{k^2}{H_0^2} \Phi_0 e^{N}),
\ee  
which again is valid on both super-Hubble and sub-Hubble scales, and we have used the initial conditions $\delta_\sigma^{(0)}=2 \delta_\chi^{(0)}=-2 \Phi_0$.

Given the solutions for the scalar field and DM perturbations we can solve for the radiation fluid perturbations.
Using the solutions \eqref{tss}, \eqref{dss}, \eqref{txs}, and \eqref{dxs} in \eqref{tr} and \eqref{dr} we have 
\bea
\theta_r^\prime - \frac{k^2}{4 H_0}e^{\frac{N}{2}} \delta_r + A \theta_r &=& \left( 1+\frac{A}{2} \right) \frac{k^2}{H_0} e^{\frac{N}{2}}\Phi_0, \label{e1}\\
\delta_r^\prime + \frac{4}{3 H_0} e^{\frac{N}{2}} \theta_r + A \delta_r &=& -A \Phi_0 - \alpha \frac{k^2}{H^2_0}e^N \Phi_0, \label{e2}
\eea
where we defined $A \equiv A_3 + A_4$ and
\be
\alpha = \frac{2}{3} \left( \frac{A_1 A_3 + 2 A_1 A_4 + A_2 A_3 +2 A_4 + A_3}{1+A_1+A_2}\right),
\ee

Differentiating \eqref{e2}, using \eqref{e1} to eliminate $\theta_r^\prime$ and \eqref{e2} to eliminate $\theta_r$ in the result, we have 
\be \label{ode}
\delta_r^{\prime \prime} + \left(2A - \frac{1}{2} \right)\delta_r^\prime+ \left( A^2 -\frac{A}{2} + \frac{k^2}{3 H_0^2} e^N \right)\delta_r= S(N),
\ee
where the source term is given by
\be \label{sources}
S(N) \equiv -\left(  A^2 -\frac{A}{2} \right) \Phi_0 - \left(  \frac{\alpha}{2}  \left(2A +1 \right) + \frac{2}{3}\left( A+2\right) \right) \frac{k^2}{H_0^2}\Phi_0 e^N .
\ee

\begin{figure}[t!]
\begin{center}
\includegraphics[scale=0.6]{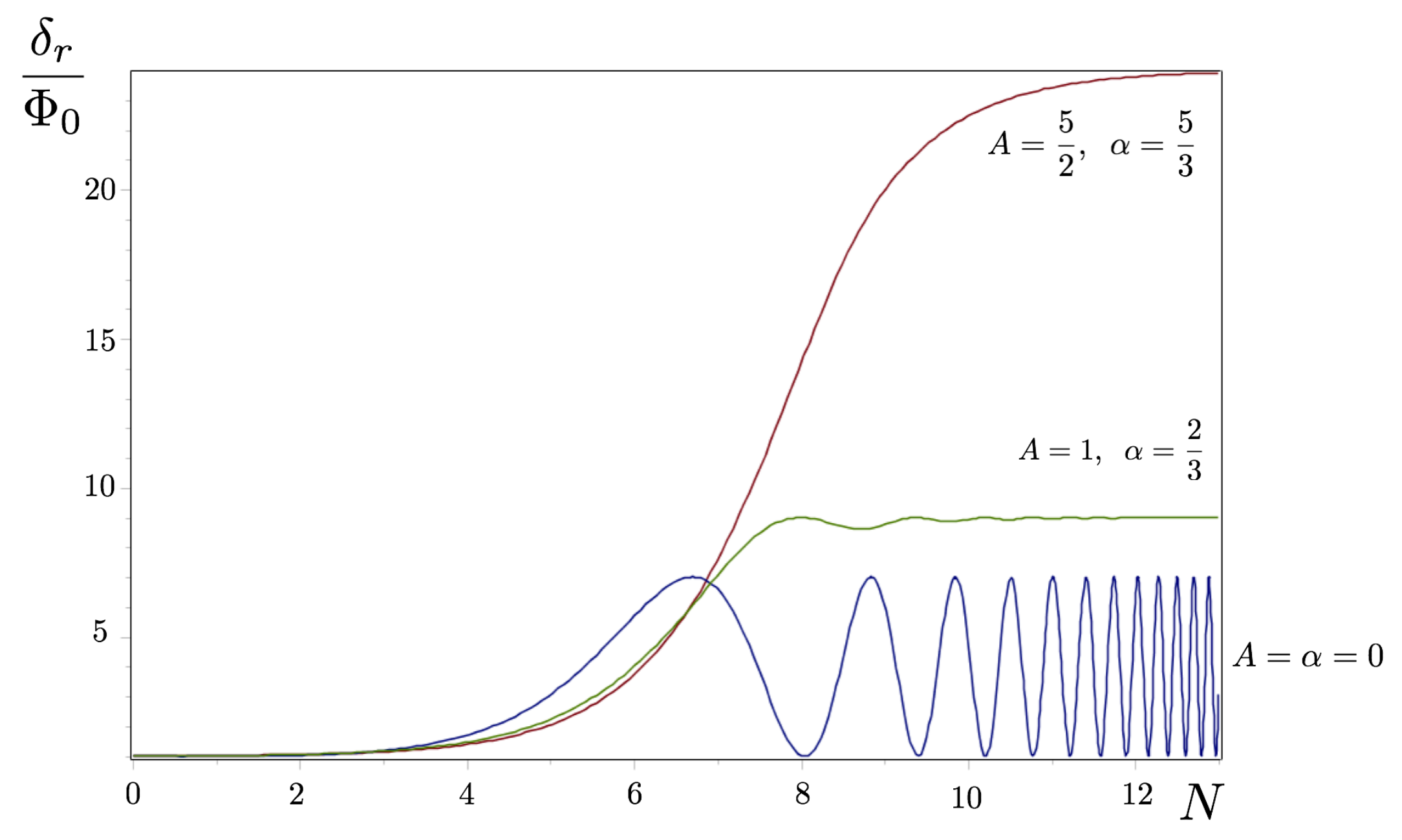}
\hspace{0.4cm}
\end{center}
\caption{Evolution of the density perturbations in radiation fluid for modes with $k/H_0 = 0.1$ and for different rates of decay and annihilation. The bottom blue curve corresponds to the evolution of radiation perturbations in the absence of scalar decay and annihilation terms in a ``matter" dominated Universe, whereas green and red curves shows the evolution with enhanced decay and annihilations. Particularly, the values of $A$ and $\alpha$ for the red curve is implied by SUSY model building.}\label{fig:deltarmdu}
\end{figure}

In the absence of decays and annihilations (corresponding to $A=\alpha=0$ above) the exact solution to \eqref{ode} can be easily found for all scales
\be
\delta_r=-4 \Phi_0 + 3 \Phi_0 \cos\left( \frac{2k}{\sqrt{3}H_0} (e^{N/2}-1)\right).
\ee
The modes are initial taken to be super-Hubble and have constant amplitude.  As they pass through the Hubble scale, they begin to 
oscillate with fixed amplitude and rapidly increasing frequency as can be seen in Figure \ref{fig:deltarmdu}.  The maximum amplitude of $| \delta_r^{max} | =  7 \Phi_0$ is reached when the lone source term in \eqref{sources} is in phase with the oscillations resulting from the homogenous solution.

There are two important differences when the scalar decay and DM annihilations are included (i.e. $A \neq 0$) as can be seen from \eqref{ode}.
Firstly, we see that if $A>1/4$ the homogeneous equation (with $S(N)=0$) becomes that of a damped oscillator.  This damping is the result of radiation being pumped into the system from decays of the scalar, as well as from DM annihilations.  The exact amount of damping depends on the relative abundances of the different fluids, the branching ratios, decay rate, and annihilation rate all given by \eqref{a3} and \eqref{a4}.  For typical initial values of radiation and DM, as well as decay rates and branching ratios as required by a successful SUSY non-thermal DM scenario, we find that typically $A>1/4$ and the oscillations in the scalar dominated phase will be over damped.  A comparison of this situation as compared with the case where annihilations and decays are absent is presented in Figure \ref{fig:deltarmdu}.   A second important effect resulting from decays and annihilations is that this provides additional source terms in \eqref{sources}, which act to boost the amplitude of the density perturbations.  As can be seen from \eqref{ode} and \eqref{sources}, unlike the $A=0$ case, the first two source terms in \eqref{sources} will lead to an immediate boost to the perturbation as it enters the Hubble radius.  The enhancement of the amplitude is again controlled by the decay and annihilation rates given by \eqref{a3} and \eqref{a4}.  In addition, although these new source terms with $A\neq 0$   provide additional enhancement, for typical values of the parameters the damping overcomes this effect before one oscillation can complete as can again be seen in Figure \ref{fig:deltarmdu}.  
 
Saturation of the radiation density perturbation at late times in the presence of decays and annihilations ($A\neq 0$) can be also understood considering the first order equations \eqref{e1} and \eqref{e2}. As we mentioned before, upon horizon entry the radiation density perturbation gets a kick and grows considerably. From \eqref{e1}, this growth began to contribute as an additional source for the velocity perturbation, causing a spatial dispersion of the radiation fluid. As the radiation velocity perturbation grows, this slows down the growth of radiation density perturbation through \eqref{e2}. Eventually, the growth in the velocity perturbation will balance the source terms in \eqref{e2} and saturate the growth in the radiation perturbation, giving that the radiation density perturbation is constant. For the full solutions of the radiation velocity and density perturbations during scalar decay we refer the reader to the Appendix C where we provide exact solutions using Green's function methods.

To summarize, in this section we have derived the  analytic solutions for the DM and radiation perturbations during the scalar dominated epoch prior to reheating\footnote{We have seen that decay of the scalar to radiation and DM is important (e.g. it changes the scaling of both radiation and DM), however the energy density of the scalar field is only reduced appreciably near the time of `reheating' $t_{rh} \sim H^{-1} \sim \Gamma_\sigma^{-1}$ as usually assumed in models of {\it instant} reheating.}. Note that the solutions we found are valid on all scales and the behavior
of perturbations in different regimes can be inferred by considering the limits $k/aH \ll 1$ or $k/aH \gg 1$. In the next section we consider the evolution of perturbations by focusing on the reheating era during which the decay term for the scalar will significantly influence the evolution of the cosmological background.
\subsubsection{Evolution of Perturbations through Reheating}   
\begin{figure}[t!]
\begin{center}
\includegraphics[scale=0.580]{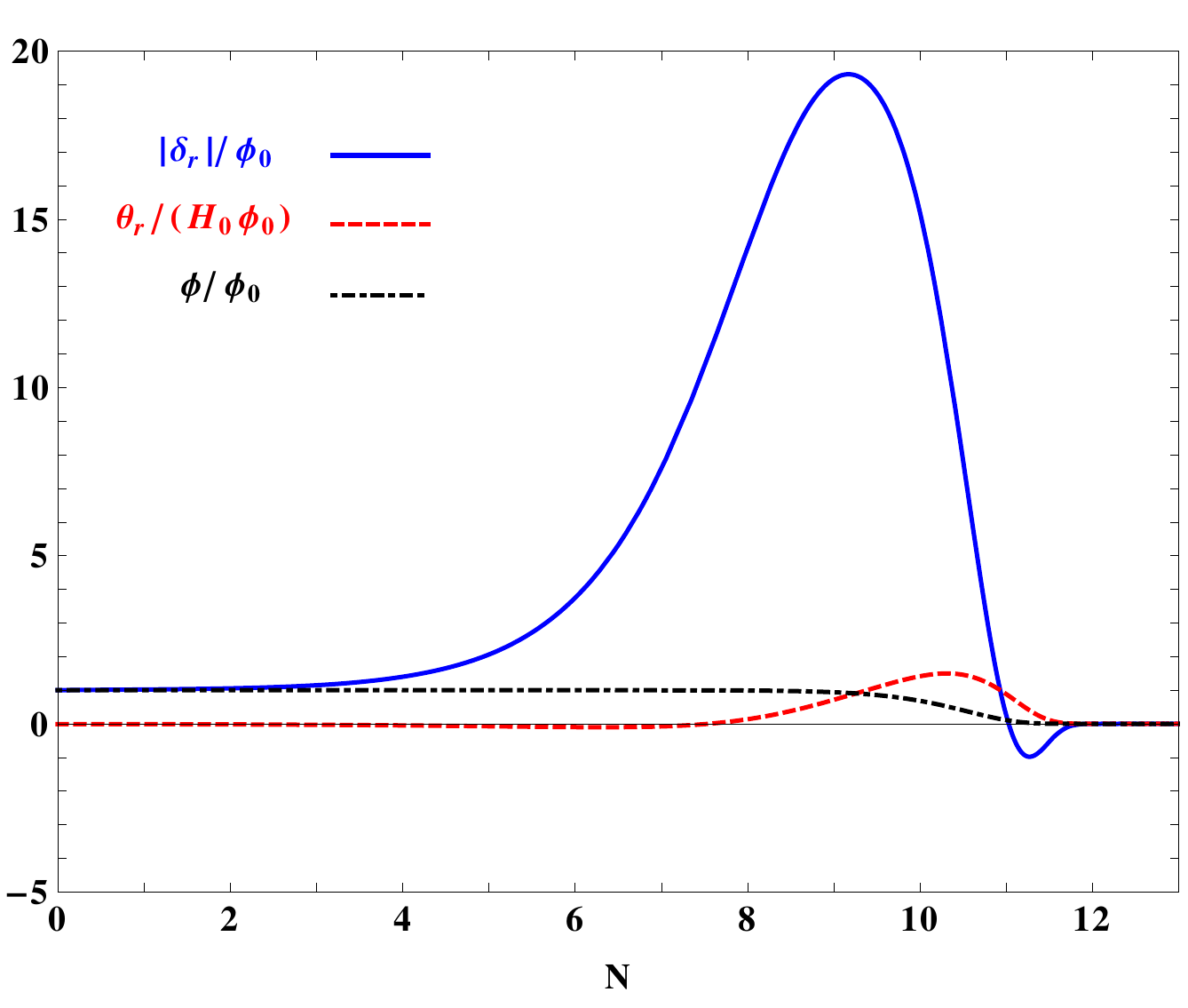}
\includegraphics[scale=0.595]{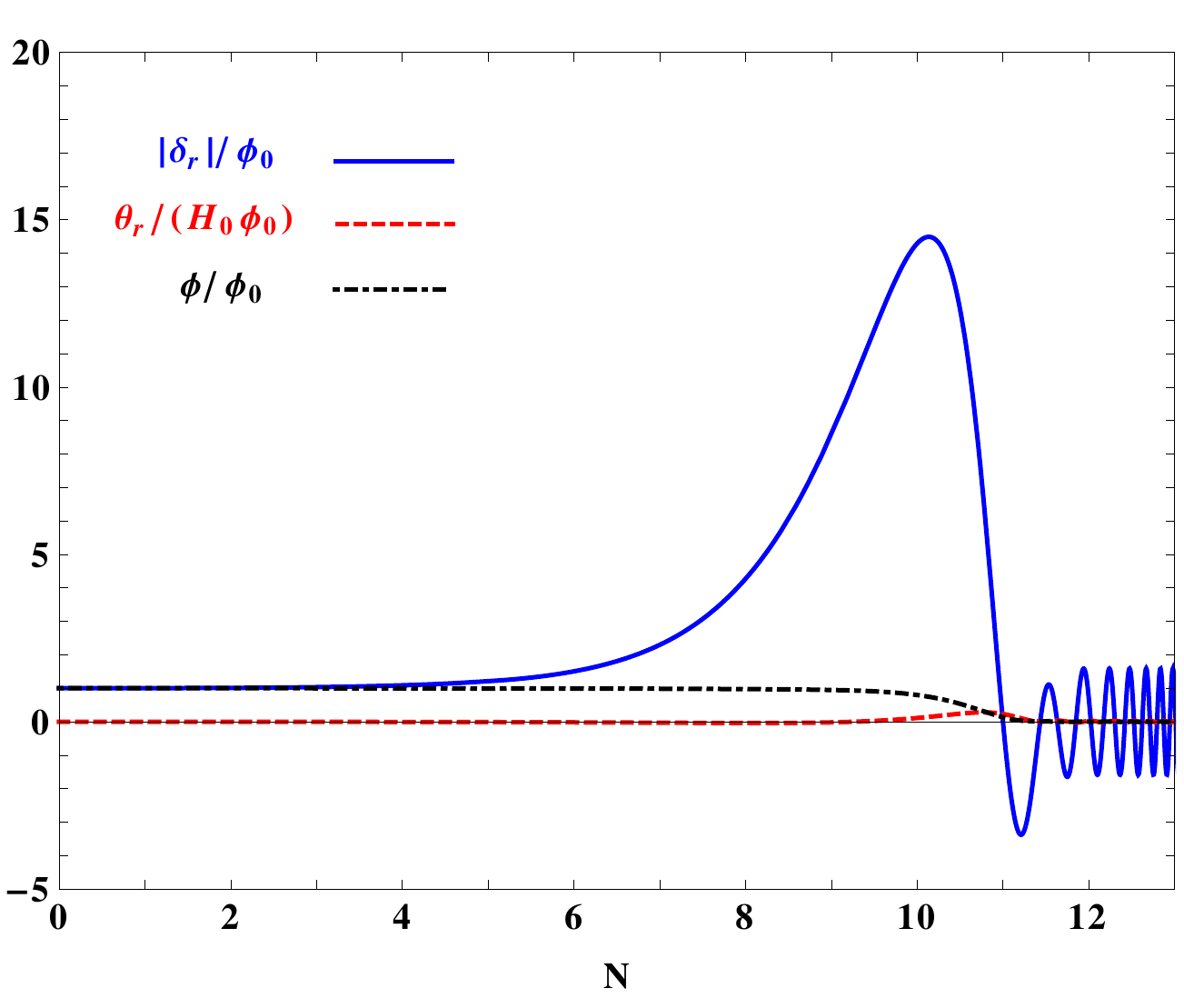}
\hspace{0.4cm}
\end{center}
\caption{Evolution of radiation density contrast and velocity perturbation (both normalized) for the modes $k/H_0=0.1$~(Left), $k/H_0=0.04$~(Right). This mode crosses the horizon at $N_h=\ln{(k/H_0)^{-2}}\simeq 4.6$~(Left), $N_h=\ln{(k/H_0)^{-2}}\simeq 6.4$~(Right). In this non-thermal cosmology the universe is effectively matter dominated until $N_{rh}\simeq 10.6$ e-foldings after which the universe becomes radiation dominated. Well after reheating, the density perturbation oscillates with an amplitude $A_{\dr}\simeq 0.0005\Phi_0$~(Left), $A_{\dr}\simeq 1.7\Phi_0$~(Right). For these modes, the ratio of the size of the comoving horizon $k_{rh}^{-1}$ at the time of reheating to the size $k^{-1}$ of the fluctuation is given by, $k/k_{rh}\simeq20$~(Left) and $k/k_{rh}\simeq8$~(Right).}\label{fig:RadP}
\end{figure}
Thus far we have neglected the effect of decays on the moduli energy density $\rhos$ and so also the effect on the Hubble expansion.
We find that this approximation will remain valid until a time near $t_d \sim H^{-1}_d \sim \Gamma^{-1}$ (or in e-folds $0<N<N_{rh}$). 
As mentioned above, this effective matter dominated phase is what allowed us to simplify the background dependent source terms in \eqref{dx}, \eqref{dr}, \eqref{tx} and \eqref{tr} (due to the scaling in \eqref{a1} -- \eqref{a4}). However, as the scalar decays become important the constant scaling is no longer valid and the evolution of these terms must be considered.  In this regime we perform the analysis numerically with our results appearing in Figures 3 and 4.  We now discuss the behavior of these solutions and their connection to the perturbation equations.

First we consider the behavior of the radiation perturbations, which is given in Figure \ref{fig:RadP}.  In the figure we show the evolution for two different modes with $k/k_{rh}=20$ and $k/k_{rh}=8$, where $k^{-1}_{rh}=1/(a_{rh}H_{rh})$ is the size of the comoving horizon at reheating. As discussed above, the radiation density perturbation gets an initial kick at Hubble radius crossing and grows considerably until it levels out due to the balance between the source terms in \eqref{dr} at around $N\simeq 9$ e-foldings. For $9<N<N_{rh}$, radiation velocity perturbations continue to grow which leads to a dispersion of the radiation density perturbations through its effect given by \eqref{dr}. Equivalently this can be understood as the importance of the friction term and sources appearing in \eqref{ode}, acting to balance each other.  The source terms lead to rapid growth of the density perturbation, but the friction term eventually saturates this growth depending on the amount of decay and annihilations.  
\begin{figure}[t!]
\begin{center}
\includegraphics[scale=0.64]{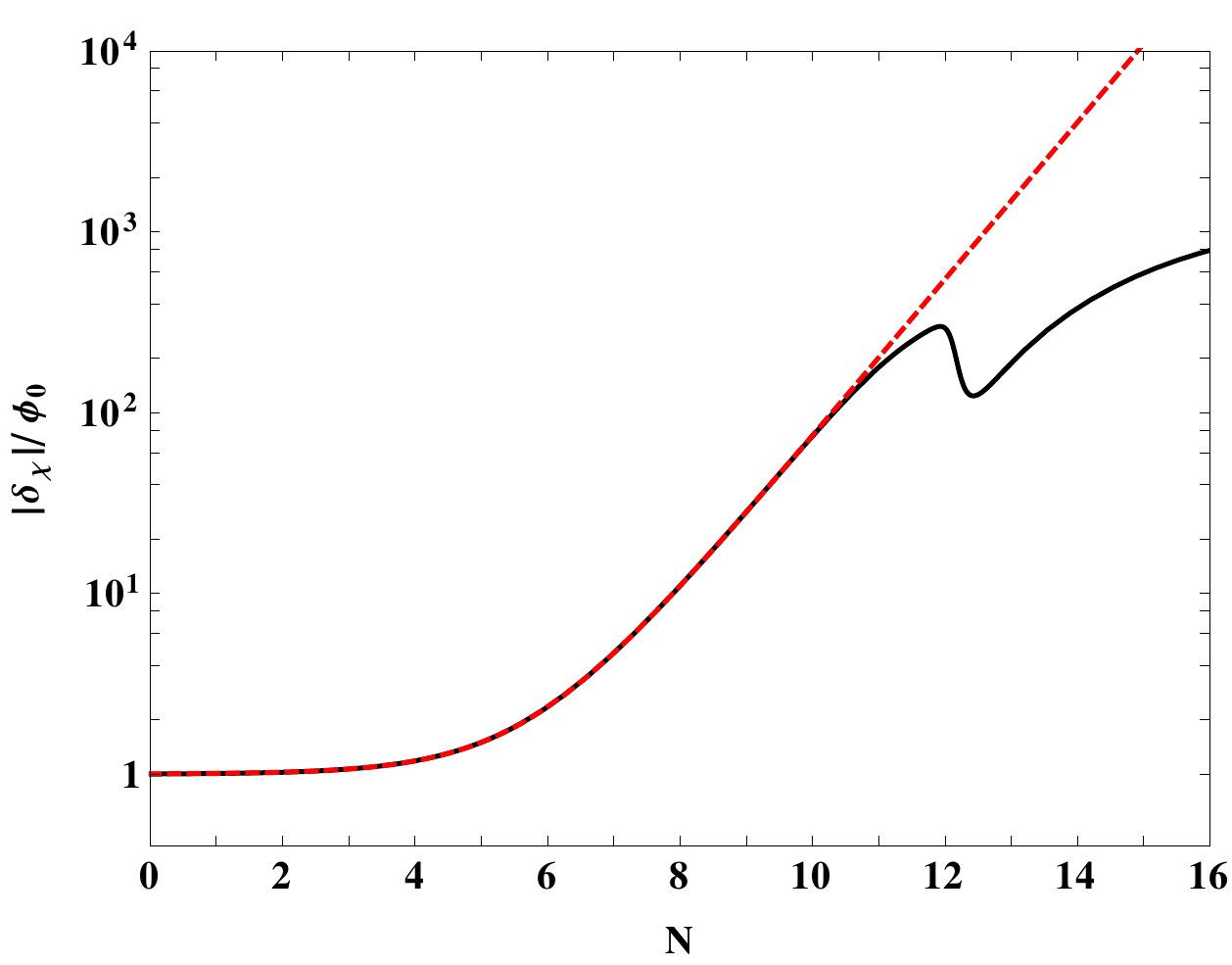}\includegraphics[scale=0.65]{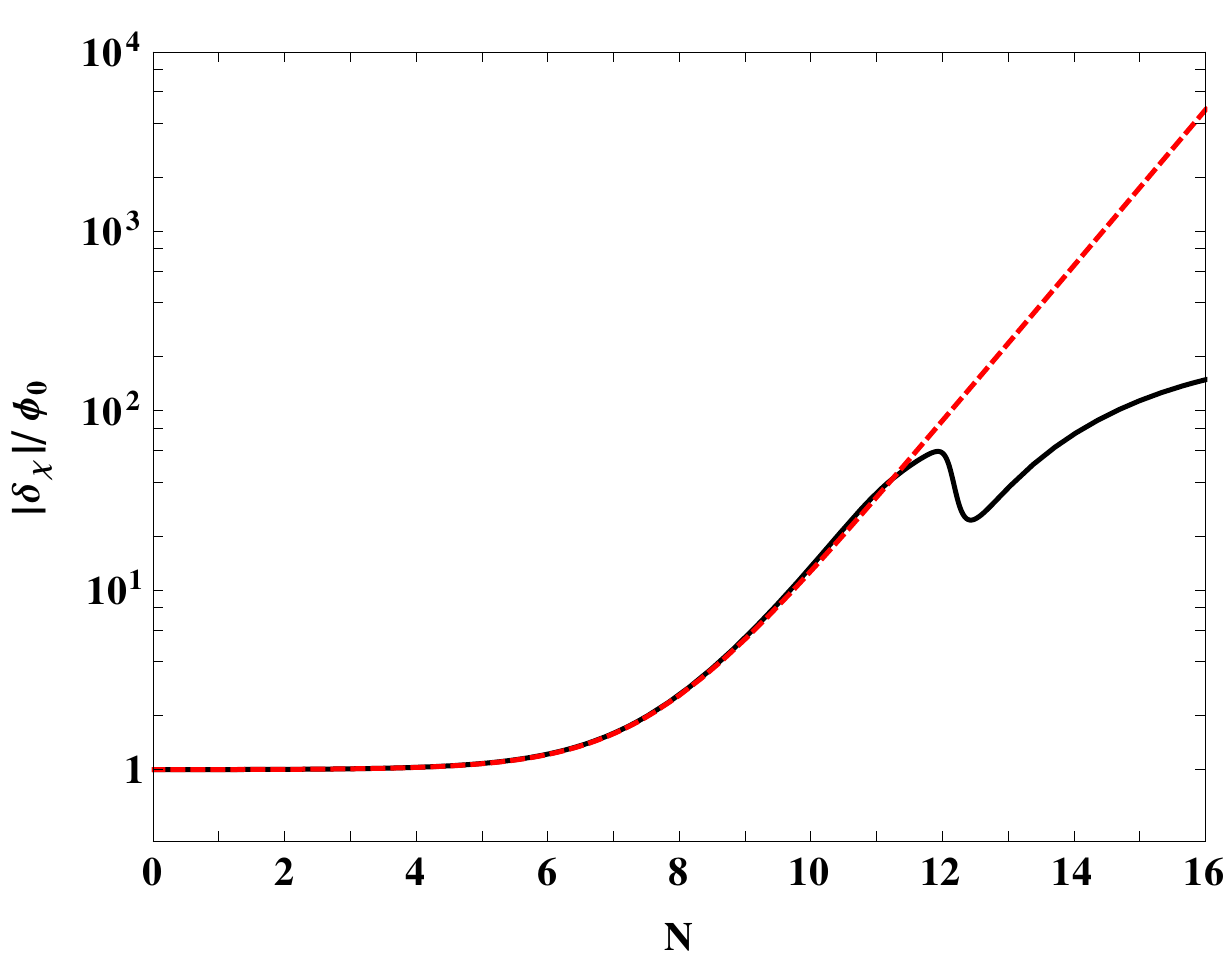}
\hspace{0.5cm}
\end{center}
\caption{Evolution of DM density contrast (normalized) in a non-thermal cosmology for modes with $k/H_0=0.1$ (Left) and $k/H_0=0.04$ (Right). As the mode enters the horizon, it  grows linearly with the scale factor $e^{N}\sim a$. We see that the solution \eqref{dxs} (red dot-dashed curve) we derived in the previous section is an excellent fit during the scalar dominated era. After the universe become radiation dominated at $N_{rh}\simeq 10.6$, the amplitude of the density contrast decreases due to rapid annihilations of DM particles. For $N\gtrsim 12$, the density contrast then begins to grow logarithmically as expected.}\label{fig:DMP}
\end{figure}

Once moduli decay becomes significant to change the expansion history at $t_d$ the moduli density then scales as $\rhos \sim e^{-2\gam/3H}$ for $N>N_{rh}\simeq10.6$, which leads to quick decay of the moduli in less than a Hubble time.  This rapid decay results in a termination of the source terms in \eqref{dr}, while the relativistic conversion of scalar particles to radiation acts to wipe out the growth prior to decay.  Thus, the only remnant of the moduli epoch is an extra suppression in the amplitude of radiation perturbations (as a consequence of the decay), which as discussed in \cite{Erickcek:2011us}  could lead to damping of dark matter perturbations on small scales if dark matter is not kinetically decoupled \cite{Loeb:2005pm}.
We find that the addition of DM annihilations to radiation does not change this conclusion.  In fact, we saw above that both annihilations and decays enter \eqref{ode} in a similar way (through the parameter $A=A_3+A_4$, see also \eqref{a3} and \eqref{a4}).  We find that the key effect of adding DM annihilations to the system is to effectively add an additional source of radiation complementary to that provided by scalar decay.  As can be seen from both Figures \ref{fig:RadP} and \ref{fig:deltarmdu}, this increases the rate of initial growth and the importance of the friction effect.  This addition also leads to a larger suppression at the time $t_d$ as the rapid decay of moduli to DM 
leads to annihilations in the super-critical case and these annihilations pump additional radiation into the system acting to further dilute the amplitude of radiation density perturbations.  

We now consider the evolution of DM perturbations.  Figure \ref{fig:DMP} shows the evolution for the DM density perturbations again for two different modes with $k/k_{rh}=20$
and $k/k_{rh}=8$.  As seen from the figure, the density contrast begins to grow linearly with the scale factor upon horizon entry until reheating $N_{rh}\simeq 10.6$. The solution \eqref{dxs} we obtained in the previous section is an excellent fit (red-dashed line) to the numerical solution for typical values of $A_1$ and $A_2$ motivated by SUSY model building. Briefly after reheating $N>N_{rh}$ the DM annihilation terms become the main source in \eqref{dx} and this along with the radiation production from decay leads to a power loss in DM density contrast. The annihilations happen in much less than a Hubble time and the resulting density contrast begins to grow logarithmically as expected in a radiation dominated universe~\cite{Brandenberger:1992dw}. In the absence of annihilations we find agreement with the analysis in \cite{Erickcek:2011us}, whereas in the case of super-critical non-thermal DM production -- when annihilations are important -- we find this acts to further reduce the strength of the perturbations following reheating.

We conclude this section by noting that the DM perturbations on scales that enter the Hubble-horizon during the early ``matter" dominated epoch can experience a significant growth. This growth might lead to formation of substructure in the form of compact mini-halos or other objects~\cite{Erickcek:2011us}, which could provide an important observational prediction of non-thermal cosmologies.
To investigate this possibility one needs to take into account cut-off scales that arise due to kinetic decoupling and free-streaming of DM candidates. We address these questions in the next two sections.

%%%%%%%%%%%%%%%%%%%%%%%%%%%%%%%%%%%%%%%%%%%%%%%%%%%%%%%

\section{Determining the Scale of the Smallest Dark Matter Structures \label{decoupling}}

\subsection{Free Streaming and Kinetic Decoupling of Dark Matter}

In the standard WIMP paradigm, while freeze-out signals the departure from {\em chemical} equilibrium it does not signal the end of WIMP interactions. Scattering processes of the form $\chi l\to \chi l$ (elastic) or $\chi l\to \chi' l'$ (inelastic) keep WIMPs in {\em kinetic} equilibrium until later times and therefore to lower temperatures~\cite{Boehm:2000gq,Chen:2001jz,Hofmann:2001bi}. 
Here $l$ and $l'$ are the light degrees of freedom in the thermal bath, while $\chi'$ is an unstable state that carries the same conserved quantum number as $\chi$. As the Universe expands and cools these scattering processes cease to be in effect and DM particles $\chi$ kinetically decouple at a temperature $T_{kd}$ when the scattering rate $\gamma$ of the DM species becomes comparable to the Hubble expansion rate, $\gamma(T_{kd})\simeq H(T_{kd})$.  

The temperature $T_{kd}$ at kinetic decoupling determines the length scale at which linear density perturbations of DM get damped, setting the scale of the smallest structures in the universe. In general, there are two important scales associated with kinetic decoupling below which the perturbations in the DM get suppressed:
\begin{enumerate}[label=\roman*.]
\item The free streaming distance of DM particles after kinetic decoupling, $k_{fs}^{-1}$,
\item The size of the comoving horizon at kinetic decoupling, $k_d^{-1}$.
\end{enumerate}

For temperatures $T<T_{kd}$, scattering of DM particles from the relativistic plasma cease to occur and WIMPs can stream from over-dense regions to under-dense regions freely, causing damping of the perturbations \cite{Bertschinger:2006nq,Green:2003un,Green:2005fa,Boyarsky:2008xj}, 
\be\label{fs}
k_{fs}^{-1}=  \int_{t_{kd}}^{t_0} \fr{\langle v \rangle}{a} dt,
\ee 
where $\langle v \rangle$ is the average velocity of DM particles after kinetic decoupling and $t_{kd}$ is the time of kinetic decoupling. However, kinetic decoupling is not an instantaneous process: coupling of   the DM fluid to acoustic oscillations in the radiation bath induces both oscillations and damping of perturbations that crosses the horizon before kinetic decoupling~\cite{Loeb:2005pm,Bertschinger:2006nq}. This leads to an additional source of damping, as so critical scale as modes with comoving wavelengths $k^{-1}<k_d^{-1}=(a_{kd}H_{kd})^{-1}$ will be damped while $k^{-1}>k_d^{-1}=(a_{kd}H_{kd})^{-1}$ continue to grow logarithmically.  We emphasize that this damping due to acoustic oscillations is most important in a universe that is dominated by the radiation bath.

\subsection{Determining the Relevant Scale}

We now consider the different possible cases for DM kinetic decoupling as compared to the reheating effects discussed in Section \ref{perts} and establish the implications for the formation of primordial substructures.
We can capture the enhanced growth of perturbations discussed in Section \ref{perts} by introducing a Gaussian cutoff into the matter power spectrum\footnote{In the simplest case of DM produced from two-body decays, we approximate the damping effects on the growth of DM perturbations by a Gaussian cut-off (See also \cite{Lin:2000qq} ).}:
\be\label{co}
\dx \to \exp\left[-\fr{k^2}{2k_{rh}^2}\fr{k_{rh}^2}{k_{cut}^2}\right] \dx(N_{rh}),
\ee
where the damping scale is given by $k_{cut}^{-1}=max(k_{fs}^{-1},k_{d}^{-1})$ -- where $k_{fs}^{-1},k_{d}^{-1}$ are the free-streaming and kinetic decoupling scales discussed above.
This expression implies that any fluctuation of size $k^{-1}$ must be larger than both of these scales to form structure.

DM particles produced from the scalar decays can thermalize with the relativistic plasma if the scattering rate is larger than the expansion rate $\gamma(T_{rh})> H(T_{rh})$. Here, we assume that the decay populates DM almost instantaneously as can be verified from Figure \ref{fig:BG}. In this case, the kinetic decoupling temperature is lower than the reheat temperature $T_{kd}<T_{rh}$ and the damping scale is typically given by $k_{cut}^{-1}={k_d^{-1}}$ as $k_{d}^{-1}>k_{fs}^{-1}$~\cite{Loeb:2005pm}. In this scenario, DM particles will lose their memory to the growth \eqref{dxs} prior to reheating and will follow the tiny oscillations in the radiation perturbation after reheating (see figure \ref{fig:RadP}).  Therefore, the growth during the scalar dominated era will be erased. This is simply because the size of the horizon at kinetic decoupling $k_d^{-1}$ is greater than the characteristic scale $k_{rh}^{-1}$ in non-thermal cosmology. The hierarchy between these scales combined with the scales of interest during the scalar dominated epoch, $k_d^{-1}>k_{rh}^{-1}>k^{-1}$ directly translate into the suppression of DM perturbations which can be  seen from \eqref{co} by noting $\dx(N_{rh})\sim (k/k_{rh})^2$ from \eqref{dxs}.

On the other hand, DM picks up a momentum when produced from heavy scalar decays after reheating and then their free streaming may erase structures on scales smaller than the free-streaming horizon $k_{fs}^{-1}$ if they don't thermalize with radiation bath. This case corresponds to $T_{kd}>T_{rh}$ where DM decouples kinetically prior reheating. In the following, we will derive the free-streaming horizon and discuss in which cases the free-streaming effect could become important. 

We denote the average momentum of DM particles produced from decays by $\langle p_{rh} \rangle$. The momentum redshifts as $a^{-1}$ assuming that there is negligible interaction to change the momentum, which is true in the case $T_{kd} > T_{rh}$: 
 \beq \label{eq:momentum}
\langle p (t) \rangle = \frac{ \langle p_{rh} \rangle a_{rh}}{a(t)}.
\eeq
 In general, the DM particles produced from decays will have a continuum spectrum and $\langle p_{rh} \rangle$ is model dependent. In the simplest case where DM particles all come from two-body decay $\sigma \to \chi \chi$, $\langle p_{rh} \rangle= \sqrt{\left(\frac{m_\sigma}{2}\right)^2-m_\chi^2}$. 

For DM particles produced from scalar decays, the free-streaming horizon in \eqref{fs} is an integration from the reheating time $t_{rh}$ till now $t_0$
\begin{eqnarray} \label{eq:freestreaming}
k_{fs}^{-1} &=& \int^{t_{0}}_{t_{rh}} \frac{\langle v \rangle}{a} dt \nonumber \\
&=&\int^1_{a_{rh}}  \left\langle \frac{p}{E} \right\rangle \frac{1}{ a^2 H} da = \int^1_{a_{rh}}  \left\langle \frac{p}{\sqrt{p^2+m_\chi^2} }\right\rangle \frac{1}{ a^2 H} da, \nonumber \\
&=& \frac{a_{rh}}{\sqrt{\Omega_R}H_0} \int^1_{a_{rh}} \left\langle  \frac{ p_{rh}}{\sqrt{p_{rh}^2a_{rh}^2+m_\chi^2a^2} }\right \rangle \frac{1}{\sqrt{1+a/a_{eq}}} da,
\end{eqnarray}
where in the second line, we changed variable from time to scale factor and used the kinematic relations $v=p/E$ and $E= p^2 +m_\chi^2$. In the third line, we used Eq.~(\ref{eq:momentum}). We also used the facts that the scale factor at matter-radiation equality could be written as $a_{eq} = \Omega_{R}/\Omega_M$ with $\Omega_R (\Omega_M)$ being the current radiation (matter) density of the Universe and 
\beq
\frac{H(a)}{H_0} = \sqrt{\Omega_R a^{-4} + \Omega_M a^{-3}}=a^{-2} \sqrt{\Omega_R}\sqrt{1+a/a_{eq}}, 
\eeq
where $H_0$ is the current Hubble rate and we neglect the dark energy contribution. 

Knowing the momentum distribution of DM after reheating, one could carry out the integration in Eq.~(\ref{eq:freestreaming}) either analytically or numerically. Below we will consider two interesting limits to obtain simple illustrative analytic results. In the case with $\langle p_{rh} \rangle \ll m_\chi$, where the DM particles produced from decays are non-relativistic, $p_{rh} \approx m_\chi v_{rh}$ with $v_{rh}$ the DM velocity after reheating, Eq.~(\ref{eq:freestreaming}) is reduced to the result in~\cite{Erickcek:2011us},
\beq
k_{fs}^{-1} \approx \frac{2 \langle v_{rh} \rangle a_{rh}}{\sqrt{\Omega_R}H_0} \left(\sinh^{-1}\sqrt{\frac{a_{eq}}{a_{rh}}}-\sinh^{-1}\sqrt{a_{eq}}\right).
\eeq
In the case with $\langle p_{rh} \rangle \gg m_\chi$, where the DM particles produced from decays are relativistic, the dominating contribution to $k_{fs}^{-1}$ is the integration over the period when DM particles are relativistic, 
\beq
k_{fs}^{-1} \approx \frac{1}{\sqrt{\Omega_R}H_0} (a_{nr} - a_{rh}),
\eeq
where we approximated the integrand in Eq.~(\ref{eq:freestreaming}) to be unity and integrated from $a_{rh}$ till $a_{nr}$ when DM is red-shifted and becomes non-relativistic with momentum of order $m_\chi$. 

If $k_{rh}/k_{fs} > 1$, the free-streaming of the DM particles will completely erase the growth of density perturbation in the scalar domination period ( See eq. \eqref{co}). In the two limiting cases we considered, the ratio $k_{rh}/k_{fs}$ is given by
\begin{eqnarray}
\frac{k_{rh}}{k_{fs}} \approx \left\{ \begin{array}{ll}
        2\langle v_{rh} \rangle  \left(\sinh^{-1}\sqrt{\frac{\sqrt{2}k_{rh}}{k_{eq}}}-\sinh^{-1}\sqrt{a_{eq}}\right), &\langle p_{rh} \rangle \ll m_\chi \\
        \frac{a_{nr}}{a_{rh}}-1 \approx \frac{\langle p_{rh} \rangle}{m_\chi}, & \langle p_{rh} \rangle \gg m_\chi . \end{array} \right. 
\end{eqnarray}
In deriving the formulas above, we used $k_{rh} a_{rh} = H(a_{rh})a_{rh}^2=H_0 \sqrt{\Omega_R}$ and $a_{eq}/a_{rh} = \sqrt{2} k_{rh}/k_{eq}$. We also chose $a_{nr} \equiv \langle p_{rh} \rangle a_{rh}/m_\chi$. To obtain more accurate numerical results in a general case, one should use Eq.~(\ref{eq:freestreaming}). Yet the approximate formulas already tell us the conditions in which case the free-streaming effect is important. If there is a large mass splitting between scalar and DM, $\langle p_{rh} \rangle \gg m_\chi$, then $k_{rh} \gg k_{fs}$ and the free-streaming effect will definitely wipe out the growth of perturbation in the scalar domination phase. Only when DM particles produced from decays are non-relativistic, $k_{rh}$ could be smaller than $k_{fs}$. Since 
\beq
\frac{k_{rh}}{k_{eq}} = 1.2 \times 10^6 \frac{T_{rh}}{1\,{\rm MeV}} \left(\frac{10.75}{g_{*,s}}\right)^{1/3} \left(\frac{g_*}{10.75}\right)^{1/2}
\eeq
and $\sinh^{-1}x$ behaves as $\log x$ for $x \gg 1$, $k_{rh}/k_{fs}$ depends weakly on $T_{rh}$ and for reheating temperatures above 10 MeV, $k_{rh}/k_{fs} < 1$ leads to $\langle v_{rh} \rangle < 0.06$~\cite{Erickcek:2011us}. This could only occur in scenarios where the scalar mass is very close to the total mass of all decays products (a situation not favored by SUSY motivated phenomenology). 
  
\section{Neutralino Dark Matter in the Moduli Scenario \label{split}}
In this section we consider SUSY neutralino DM in the Split SUSY / moduli framework, which provides an explicit realization of the non-thermal histories discussed above.
We will be interested in wino or higgsino DM and we begin by summarizing the picture of kinetic decoupling.

 In the moduli scenario, immediately after the moduli decay at about $T_{rh}$, the produced DM particles would have a energy distribution which peaks at high energy and most of them are relativistic. Through scattering with SM particles, $e^-, \nu_e, \nu_\mu, \nu_\tau$ in the thermal bath, they will deposit energy into radiation. Because the scattering rate of either wino or higgsino is large enough (compared to Hubble), the DM particles will thermalize with the radiation quickly. When the temperature decreases and the DM particles become non-relativistic, the rate of the thermal scattering drops and eventually the DM particles would be kinetically decoupled from radiation. One key quantity that will determine whether thermalization happens or not is $\gamma/H$ where $\gamma$ is the scattering rate of DM particles off radiation. In this section, we will demonstrate thermalization indeed happens for either wino or higgsino DM by computing their scattering rates $\gamma/H$'s.

Light wino DM with mass of about a few hundred GeV fits very nicely into the moduli scenario~\cite{Moroi:1999zb}. However, indirect detection searching for excesses in the photon spectrum of our galactic center have already put strong constraints\footnote{Weaker constraints on the reheat temperature were found in \cite{Easther:2013nga}, but this analysis only took into account FERMI observations of dwarf spheroidal galaxies. Reach of future dwarf spheroidal galaxies observations has been studied in \cite{Bhattacherjee:2014dya}.} on the wino as the only component of DM~\cite{Cohen:2013ama, Fan:2013faa}. In particular, $T_{rh} > 1$ GeV in the moduli scenario even if the wino is only one component of DM~\cite{Fan:2013faa}~\footnote{The constraint could be relaxed if the branching fraction of moduli decaying to winos are suppressed as in the branching scenario~\cite{Allahverdi:2013noa}.}. The cosmological history of winos after reheating has already been worked out in detail in Ref.~\cite{Arcadi:2011ev}. It is demonstrated that winos lose energy efficiently after production through the inelastic process $\tilde{W}^0 + e^{\pm} \to  \tilde{W}^\pm + \nu_e$ and thermalize with the radiation almost instantaneously. At low temperature $T_{kd} \approx 10$ MeV, the wino DM will kinetically decouple from the thermal bath. Notice that $T_{kd}$ is almost independent of the wino mass and is mostly set by the mass splitting between the charged and neutral components of wino DM, which is about $\Delta m \approx 160$ MeV at the two-loop level~\cite{Ibe:2012sx}. This is because at low temperature, the scattering rate will be suppressed by the Boltzmann factor exp$(-\Delta m/T)$. 

Now we turn to the higgsino DM scenario. When DM is mostly higgsinos, or in other words, $\mu < M_1, M_2$, unlike the wino case, the tree-level mass splitting is only suppressed by one power of the larger mass scale $M_1$ or $M_2$,
\beq
\Delta m^{\tilde H} \approx \frac{m_Z^2}{2M_1}c_W^2\left(1-\sin 2\beta\right)+\frac{m_Z^2}{2M_2}s_W^2\left(1+\sin 2\beta\right) \approx  0.5 \, {\rm GeV} \frac{4.5 \, {\rm TeV}}{M_s}, \label{eq:splitting}
\eeq
where in the second step, we assume that $\tan \beta \gg 1$ and $M_2 =2M_1=M_s$. Traditionally one could diagonalize the neutralino/chargino mass matrices and expand the formulas to obtain the result above. However, this could also be understood easily from an effective operator analysis. Integrating out a heavy bino or wino at tree-level, one gets dimension-five operators such as 
\beq
\frac{g^{\prime 2}}{M_1}H_u^\dagger \tilde{H}_uH_d^\dagger \tilde{H}_d, \quad \frac{g^{2}}{M_2}H_u^\dagger\sigma^a \tilde{H}_u  H_d^\dagger\sigma^a \tilde{H}_d
\eeq
where $\sigma^a$ with $a=1,2,3$ are the three $SU(2)_w$ generators. $g$ and $g^\prime$ are the SM $SU(2)_w$ and $U(1)_B$ gauge couplings correspondingly. The operators above will lead to a charged/neutral mass splitting after electroweak symmetry breaking. Notice that the operators above also lead to an effective coupling between Higgsinos and $Z$ gauge boson.

In the moduli scenario, the relic abundance of higgsino DM is estimated to be, assuming order one branching fraction of moduli decaying to higgsinos,
\bea
\Omega_{\tilde{H}}h^2 &=& 0.12 \frac{\langle\sigma v\rangle_{th}}{\langle\sigma v\rangle_{\tilde{H}\tilde{H}\to ZZ,WW}}\frac{T_{f}}{T_{rh}}, \nonumber \\
T_{rh}&\approx & 0.4 \, {\rm GeV}\left(\frac{\mu}{200\,{\rm GeV}}\right)^3 \left( \frac{0.12}{\Omega_{\tilde{H}}h^2}\right),
\eea 
where we took $\langle\sigma v\rangle_{th} = 3 \times 10^{-26}$ cm$^3$/s and the thermal freeze-out temperature $T_{f} \approx \mu/20$. In deriving the second line, we used $\langle\sigma v\rangle_{\tilde{H}\tilde{H}\to ZZ,WW} \approx g^4/(512\pi \mu^2)(21+3\tan^2\theta_W+11\tan^4\theta_W)$~\cite{ArkaniHamed:2006mb}.
 
\begin{figure}[h]\begin{center}
\includegraphics[width=0.45\textwidth]{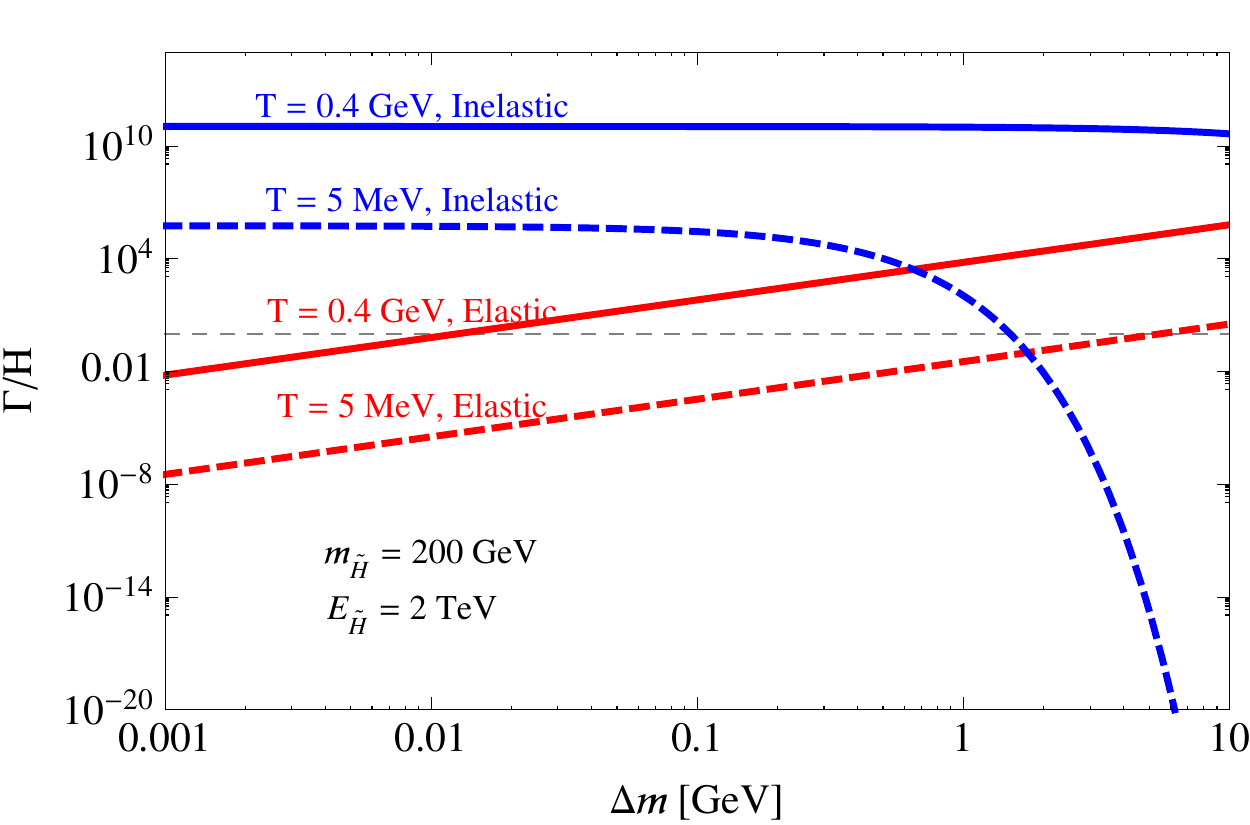} \quad \includegraphics[width=0.45\textwidth]{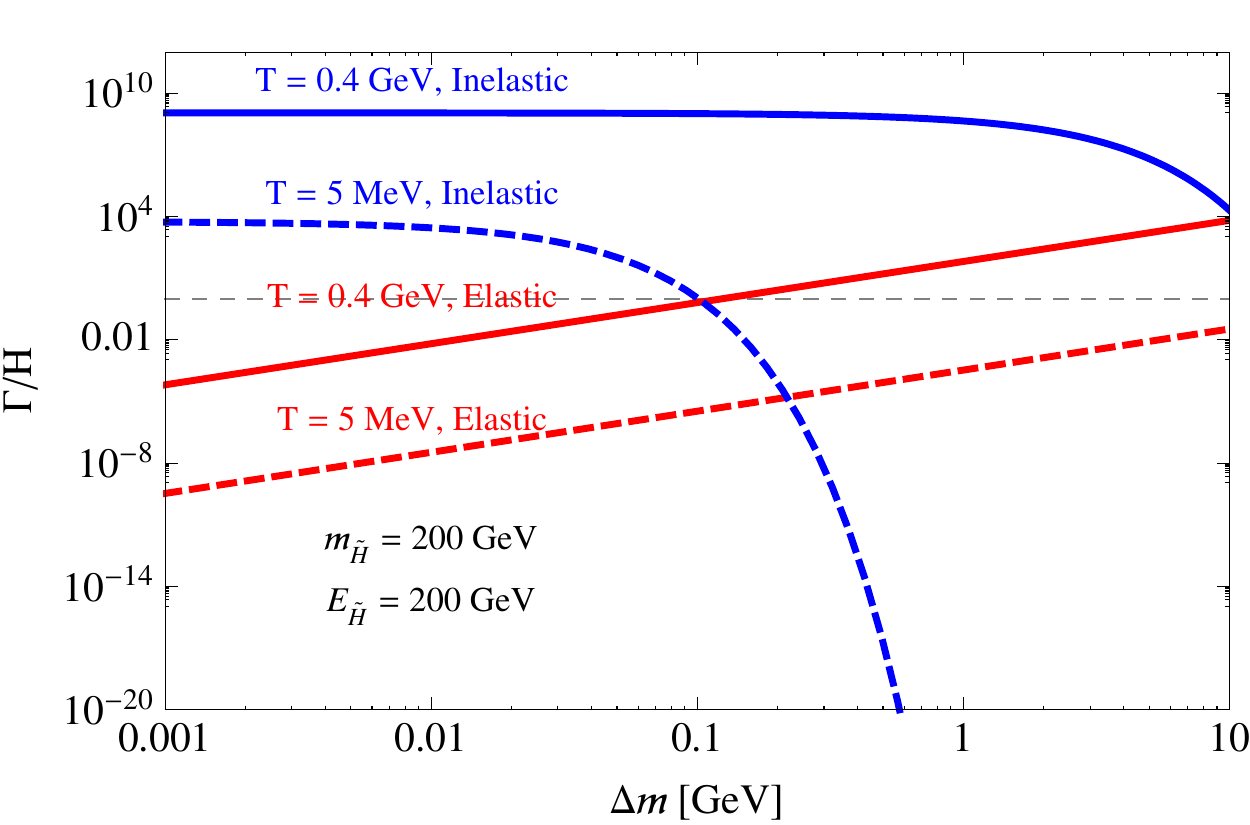}
\end{center}
\caption{Elastic/inelastic scattering rates per Hubble (red/blue curves) as a function of the mass splittings fixing $\mu = 200$ GeV. Left: energy of DM produced from decay is fixed to be 2 TeV. Right: energy of DM produced from decays is fixed to be 200 GeV. The solid curves correspond to the temperature of the thermal bath to be about $T_{rh} = 0.4$ GeV; the dashed curves correspond to a much lower temperature 5 MeV. The gray dashed lines corresponds to $\gamma/H =1$. }
\label{fig:higgsino}
\end{figure}

The elastic scattering rate of higgsino DM per expansion rate is~\cite{Hisano:2000dz} 
\beq
\frac{\gamma_{el}}{H}&=&\frac{45\sqrt{5}}{16 \pi^{9/2}}\frac{1}{\sqrt{g_*(T)}} \frac{g^4}{m_W^4} c_{\tilde{H}\tilde{H}Z}^2 (1-s_W^2+2s_W^4) \frac{m_{p} E^2T^3}{\mu^2}, \\
{\rm where} \quad c_{\tilde{H}\tilde{H}Z}&=&\frac{m_Z^2}{2 \mu} \left(\frac{s_W^2}{M_1}+\frac{c_W^2}{M_2}\right)\cos(2\beta) \nonumber \\
&=&\frac{\Delta m}{\mu} \frac{\left(s_W^2+c_W^2/r\right)\cos (2\beta)}{(c_W^2+s_W^2/r)+(s_W^2/r-c_W^2)\sin(2\beta)},
\eea 
where $E$ is the energy of the DM particles, $T$ is the temperature of the radiation bath and $r=M_2/M_1$. One could see that the elastic scattering rate per Hubble scales as $(\Delta m)^2$ and increases when $\Delta m$ increases. 
The inelastic scattering of higgsino DM per Hubble is 
\beq
\frac{\gamma_{in}}{H}=\frac{6\sqrt{5}}{ \pi^{3/2}}\frac{1}{\sqrt{g_*(T)}} \frac{g^4}{m_W^4} m_{p} ET^2 e^{-\frac{\mu \Delta m}{2 ET}} \left(\frac{\Delta m}{\mu}+6 \frac{ET}{\mu^2}\right),
\eeq
where $e^{-\frac{\mu \Delta m}{2 ET}}$ is the Boltzmann factor. Thus inelastic scattering would be more efficient at small $\Delta m$. In Fig.~\ref{fig:higgsino}, we demonstrated the higgsino elastic/inelastic scattering rates per Hubble as a function of the mass splittings for different choices of the DM energy and thermal bath's temperature. From Fig.~\ref{fig:higgsino}, the inelastic scattering rate always dominates over elastic scattering rate at the reheating temperature for the whole range of mass splitting. It is also much larger than Hubble rate and thus the higgsino DM particles would quickly thermalize with the radiation immediately after reheating. When the temperature drops, $\gamma/H$ decreases and eventually higgsinos decouple kinetically. 

%%%%%%%%%%%%%%%%%%%%%%%%%%%%%%%%%%%%%%%%%%%%%%%%%%%%%%%
%%%%%%%%%%%%%%%%%%%%%%%%%%%%%%%%%%%%%%%%%%%%%%%%%%%%%%
In summary, we find that for neutralino dark matter with a non-thermal history that any enhancement of dark matter perturbations arising from the moduli epoch are washed out 
by kinetic decoupling effects following reheating.  

\section{Conclusions \label{conclude}}
In this paper we have investigated the result of a matter (moduli) dominated phase prior to BBN on the growth of cosmological perturbations.
We have seen that matter and radiation perturbations grow during this epoch, with dark matter perturbations being enhanced and radiation perturbations first growing and then eventually oscillating with 
an amplitude suppressed relative to the usual thermal case.  We saw that this suppression arises at the peak of scalar decay when most of the radiation is created by both dark matter annihilations and decays of the moduli.  In agreement with \cite{Erickcek:2011us} we find that the matter perturbations remain enhanced following reheating (with the growth inherited from the growth of scalar fluctuations during moduli domination), and we showed that this remains true even in the presence of dark matter annihilations to radiation.  Whereas the suppression of the radiation perturbations can lead to damping of dark matter perturbations on small scales if dark matter is not kinetically decoupled \cite{Loeb:2005pm} -- again in agreement with \cite{Erickcek:2011us}.

However, we have also seen that for non-thermal models motivated by BSM physics -- such as those motivated by moduli in the presence of SUSY breaking with a split spectrum -- that these effects are lost since the kinetic decoupling temperature of neutralinos is typically below that of reheating.  This is discouraging for establishing new signatures for the dark ages following inflation, because it means that the matter power spectrum (and enhanced small scale structure such as compact mini-halos) can not be used to distinguish a non-thermal history from the standard thermal case.  Moreover, because the effects of the moduli decays to dark matter are local (sub-horizon) and dark matter is subdominant before, during, and after the decays, we should not expect any associated CMB signatures (e.g. from dark matter annihilations). That is, as far as observations are concerned, the cosmic dark ages remain elusive.  

There are some exceptions to this conclusion.  As discussed in \cite{Erickcek:2011us}, if dark matter is produced thermally after reheating (and kinetic decoupling and free-streaming effects are not important) the suppression of the radiation perturbations resulting from the non-thermal phase can lead to an erasure of dark matter structure (establishing a cutoff in the matter power spectrum).  For typical SUSY WIMP models this seems to present model building challenges given the need for a large reheat temperature, and so the matter dominated phase would be short or even comparable to that of a thermal history.    Another possibility
is if dark matter is produced non-thermally but with a mass comparable to the moduli mass (so it is not relativistic at production).  This possibility again seems rather exotic from a SUSY model building viewpoint, particularly in models with moduli masses associated with PeV-scale symmetry breaking. 

Taking a more optimistic view, our results suggest the robustness of typical non-thermal histories (or early matter dominated phases) prior to BBN and provide further evidence that such models offer a realistic alternative to the standard thermal WIMP paradigm.
 
\section*{Acknowledgements}
We would like to thank Niayesh Afshordi, Dan Hooper, Doddy Marsh, Kris Sigurdson, and Kuver Sinha for useful discussions.  We are especially grateful to Adrienne Erickcek for comments on the first version of this paper, which lead to an improved discussion of our results. The work of SW and OO is supported in part by NASA Astrophysics Theory Grant NNH12ZDA001N, and DOE grant DE-FG02-85ER40237. SW would also like to thank the DAMTP, Cambridge University and the Mitchell Institute for Fundamental Physics and Astronomy for hospitality. The research of SW was also supported in part by the National Science Foundation under Grant No. NSF PHY11-25915.

\subsection*{Appendix A: Derivation of Perturbation Equations}

We take the scalar field, radiation and the DM as perfect fluids with energy-momentum tensors
\be\label{SEM}
T^{\mu}_{~\nu}=
(\rho+p)u^{\mu}u_{\nu} + \delta^\mu_\nu p,
\ee
where four-velocities in the rest frame of each fluid is $u^\mu=(1,\vec{0})$. In this Appendix, we will work with the background metric $g_{\mu\nu}=diag(-1,a^2,a^2,a^2)$ and use cosmic time $t$ to be proper time. In the reheating model we are considering, there are energy transfer between different fluids which can be captured in a covariant manner by writing,
\be\label{EM}
\nabla_{\mu}T^{\mu{(\alpha)}}_{~\nu}= Q^{(\alpha)}_{\nu}+Y^{(\alpha)}_{\nu},
\ee 
where $Q^{(\alpha)}_{\nu}$ denotes the energy transfer due to scalar decay and $Y^{(\alpha)}_{\nu}$ stands for the transfer of energy due to annihilations. Note that the total energy conservation $\sum_\alpha \nabla_{\mu}T^{\mu{(\alpha)}}_{~\nu}=0$ implies the constraint 
\be
\sum_\alpha Q^{(\alpha)}_{\nu}+Y^{(\alpha)}_{\nu}=0.
\ee 
The covariant form of energy transfer terms on the R.H.S of \eqref{EM} can be written as,
\bea
\label{Td}Q^{(\alpha)}_{\nu} &=& \gam ^{(\alpha)} T_{\mu\nu}^{(\sgm)} u^\mu_{(\sgm)},\\
\label{Tan}Y^{(\alpha)}_{\nu}&=& \frac{\langle \sgm v \rangle^{(\alpha)}}{m_\chi}
\left[\rhom^2-\rho_{\chi,eq}^2\right]
u_{\nu}^{(\chi)},
\eea
from which one can easily obtain,
\be\label{BGA}
\dot{\rho}_{(\alpha)}+3H\ra(1+\wa)=-\gam ^{(\alpha)}\rhos+ \fr{\langle \sgm v \rangle^{(\alpha)}}{m_\chi}[\rhom^2-\rho_{\chi,eq}^2].
\ee
Then, the fluid evolution equations \eqref{BGE1}-\eqref{BGE3} can be recast using
the following conventions for the decay rate and annihilation cross section between scalar,radiation and DM fluids: $\langle \sgm v \rangle^{(\sgm)}=0,~\langle \sgm v \rangle^{(\chi)}=-\langle \sgm v \rangle,~\langle \sgm v \rangle^{(r)}=\langle \sgm v \rangle$ and $\gam ^{(\sgm)}=\gam,~\gam ^{(\chi)}=-B_\chi\gam, ~\gam ^{(r)}=-(1-B_\chi)\gam$. We write the perturbed metric in the longitudinal gauge as 
\be
ds^2=-\left(1+2 \Phi \right) dt^2 + a(t)^2\left(1-2 \Psi \right) \delta_{ij} dx^i dx^j.
\ee
In the absence of anisotropic stress, from \eqref{SEM} we write perturbed energy-momentum tensor;
\be
\delta T^{\mu(\alpha)}_{~\nu} = \left(\begin{array}{cc}-\delta \rho_{(\alpha)} &  \ra\left(1+\wa\right) \partial_i v_{(\alpha)}  \\ - g^{ij}\ra \left(1 + \wa \right) \partial_j v_{(\alpha)} & \delta^i_j~ \ca^2 \delta \rho_{(\alpha)} \end{array}\right),
\ee 
where we have used $\pa=\wa\ra$, $\delta p_{(\alpha)}= c_{(\alpha)}^{2} \delta\rho_{(\alpha)}$ and $v_{(\alpha)}$ denotes the longitudinal  part of the spatial velocity perturbation of each fluid, $\delta u_{j(\alpha)}=\partial_j v_{(\alpha)}$. To first order in scalar perturbations, components of energy transfer terms in \eqref{Td} and \eqref{Tan} reads
\bea\label{RHS}
\nn\delta Q^{(\alpha)}_{0}&=& \gam ^{(\alpha)} 
\left[\delta\rhos +\rhos\Phi\right],\\
\nn\delta Y^{(\alpha)}_{0}&=& -\fr{\langle \sgm v \rangle^{(\alpha)}}{m_\chi}\left(2\rhom\delta\rhom -
2\rho_{\chi,eq}\delta \rho_{\chi,eq}+[\rhom^2-\rho_{\chi,eq}^2]\Phi\right),\\
\delta Q^{(\alpha)}_{j}&=& -\gam ^{(\alpha)}\rhos\partial_j v_{(\sgm)},\\
\nn\delta Y^{(\alpha)}_{j}&=& \fr{\langle \sgm v \rangle^{(\alpha)}}{m_\chi}[\rhom^2-\rho_{\chi,eq}^2]\partial_j v_{(\chi)},
\eea 
where we used the fact that $\delta u_{(\alpha)}^{0}=-\Phi$. Finally, we can obtain evolution equations for the density and velocity perturbations using perturbed stress-energy conservation equation \eqref{EM} with \eqref{RHS}, setting $\Psi=\Phi$ in the absence of anisotropic stress, for temporal and spatial components we have,
\bea
\nn \dot{\delta}_{(\alpha)}+3H(\ca^2-\wa)\da +(1+\wa)\left(\fr{\ta}{a}-3\dot{\Phi}\right)&=&-\gam ^{(\alpha)}\fr{\rhos}{\ra}\left[\ds - \da + \Phi\right]\\ \nn &+&\fr{\langle \sgm v \rangle^{(\alpha)}}{m_\chi \ra}\left(\rhom^2[ 2\dx-\da +\Phi]-\rho_{\chi,eq}^2[2\delta_{\chi,eq}
-\da+\Phi]\right),\\
\nn \dot{\theta}_{(\alpha)}+H\ta +\fr{\ca^2}{1+\wa}\fr{\nabla^2\da}{a} -3H\wa\ta +\fr{\nabla^2\Phi}{a} &=& -\gam ^{(\alpha)}\fr{\rhos}{\ra}\left[\fr{\ts}{1+\wa}-\ta\right]\\ \nn &+& \fr{\langle \sgm v \rangle^{(\alpha)}}{m_\chi \ra}[\rhom^2-\rho_{\chi,eq}^2]\left(\fr{\tx}{1+\wa}-\ta\right),
 \eea
where we defined $\da\equiv\delta\rho_{(\alpha)}/\ra$, $\theta_{(\alpha)}\equiv a^{-1}\nabla^{2}v_{(\alpha)}$ and used the background fluid equations \eqref{BGA}.

 \subsection*{Appendix B: Adiabatic Initial Conditions}
Before we start our discussion on adiabatic initial conditions, we would like to briefly review isocurvature perturbations and point out the non-existence of isocurvature modes in our reheating model after the decay of modulus. 

On super-horizon scales, one can define a gauge invariant curvature perturbation for each species $\alpha$ of the universe, which is conserved in the adiabatic limit when the Hubble expansion is dominated by a single species \cite{Bardeen:1983qw} 
\be
\zeta_{(\alpha)}=-\Psi - H\fr{\delta\rho_{(\alpha)}}{\dot{\rho}_{(\alpha)}},
\ee
from which one can find the total curvature perturbation as a weighted sum of $\zeta_{(\alpha)}$;
\be
\zeta = \fr{\sum_{\alpha}(\ra+\pa)\zeta_{(\alpha)}}{\sum_{\alpha}(\ra+\pa)}.
\ee
The definition of isocurvature perturbation between two species is given by
\be
S_{\alpha\beta}=3(\zeta_{(\alpha)}-\zeta_{(\beta)}).
\ee

During inflation, in the model we are considering, the mass of the modulus at the high energy minima  satisfies, 
\be
m_{\sgm}^2 \lesssim H_{inf}^2,
\ee
with an average amplitude of long wavelength fluctuations \cite{Linde:2005ht};
\be
\delta\sgm \sim H_{inf}/2\pi.
\ee
The existence of such a mode can lead to isocurvature perturbations: while radiation and matter created from inflationary reheating carries the inflaton's fluctuation $\zeta_{inf}$, those produced from the moduli decay will inherit $\delta\sgm$. However, if the modulus dominates the energy density of the universe, all existing isocurvature modes will be washed out as shown in \cite{Iliesiu:2013rqa} recently. Therefore, consistent with CMB anisotropy probes, we will consider adiabatic perturbations in our reheating model. We will consider the issue of adiabatic modes in the presence of moduli decay and DM annihilations in some detail below. 

In the case of energy transfer between the constituents of the universe, it has been showed that perturbation equations allow for an adiabatic solution in the long wavelength limit with
\be
 \fr{\delta\rho_{\alpha}}{\dot{\rho}_{\alpha}}=\fr{\delta\rho_{\beta}}{\dot{\rho}_{\beta}},
\ee
if the total intrinsic non-adiabatic energy transfer perturbation of each individual species $\delta Q^{(\alpha)}_T=\delta Q^{(\alpha)}_{0}+\delta Y^{(\alpha)}_{0}$ vanishes~\cite{Malik:2004tf}. For the reheating model we consider, these are given by
\bea
\label{Qs}\delta Q^{(\sgm)}_T &=& 0,\\
\label{Qx}\delta Q^{(\chi)}_T &=& -B_\chi \fr{\gam}{3H}\dot{\rhos} S_{\sgm\chi},\\
\label{Qr}\delta Q^{(r)}_T &=& -(1-B_\chi)\fr{\gam}{3H}\dot{\rhos} S_{\sgm r} -\fr{2\langle \sgm v \rangle^{(r)}}{3m_\chi H} \rhom \dot{\rho}_\chi S_{\chi r},
\eea
where again we set $\rho_{\chi,eq}=0$ in the era we consider the fluid perturbation equations. Therefore, from \eqref{Qs}-\eqref{Qr}, we see that an adiabatic mode with $\zeta=constant$, $\Psi=constant$ and $S_{\alpha\beta}=0$ \cite{Malik:2004tf} exist on large-scales where we ignored the decaying mode of gravitational potential $\Psi$.  
%%%%%%%%%%%%%%%%%%%%%%%% Begin Appendix %%%%%%%%%
%%%%%%%%%%%%%%%%%%%%%%%%%%%%%%%%%%%%%%%%%%%%%%%%%
\appendix

\subsection*{Appendix C: Solution for Radiation Density Perturbation During Scalar Domination}\label{appendixC}
We are interested in finding the solution for the density perturbation during scalar domination.
The equation of motion derived in the text is
\be 
\delta_r^{\prime \prime} + \left(2A - \frac{1}{2} \right)\delta_r^\prime+ \left( A^2 -\frac{A}{2} + \frac{k^2}{3 H_0^2} e^N \right)\delta_r= S(N),
\ee
where the source term is given by
\be 
S(N) \equiv -\left(  A^2 -\frac{A}{2} \right) \Phi_0 - \left(  \frac{\alpha}{2} \left(2A +1 \right) + \frac{2}{3}\left( A+2\right) \right) \frac{k^2}{H_0^2}e^N \Phi_0.
\ee

The homogeneous solution is
\bea
\delta_r^{(h)}
&=&c_1 \, e^{-AN} \sin\left(\frac{2k}{\sqrt{3}H_0} e^{\frac{N}{2}} \right) + c_2 \, e^{-AN} \cos\left(\frac{2k}{\sqrt{3}H_0} e^{\frac{N}{2}} \right)
\eea
From this we can construct the Green's function
\bea
G(N,\tN) &=& \frac{s_1(N) s_2(\tN)-s_1(\tN) s_2(N) }{ s^\prime_1(\tN) s_2(\tN)-s_1(\tN) s^\prime_2(\tN) } \nonumber \\
&=& \frac{\sqrt{3}H_0}{k} e^{(A-\frac{1}{2})\tN-AN} \sin\left[\frac{2k}{\sqrt{3} H_0} \left(e^{\frac{N}{2}}-e^{\frac{\tN}{2}} \right) \right]
\eea
and so the full solution is
\be
\delta_r = \delta_r^{(h)} + \int_0^N G(N,\tN)  S(\tN) d\tN,
\ee
The integral gives a negligible contribution for super-Hubble modes initially, and so using the initial conditions $\delta_r(N=0)= \delta_r^{(0)}=-\Phi_0$ and 
$\delta^\prime_r(N=0)=0$ for $k<aH$ we find 
 $c_1=\sqrt{3}A H_0  \delta_r^{(0)} / k$ and $c_2=(1-2A)\delta_r^{(0)}$.
The full solution is then 
\bea \label{fullsoln}
\delta_r&=& c_1 e^{-AN} \sin\left(\frac{2k}{\sqrt{3}H_0} e^{\frac{N}{2}} \right)+c_2 \,e^{-AN} \cos\left(\frac{2k}{\sqrt{3}H_0} e^{\frac{N}{2}} \right) + \delta_r^{(p)},
\eea
where the particular solution is given by
\bea
\delta_r^{(p)} &=&  \int_0^N G(N,\tN)  S(\tN) d\tN, \nonumber \\
&=& \frac{\sqrt{3}H_0}{k}e^{-AN} \Phi_0 \int_0^N  e^{(A-\frac{1}{2})\tN} \sin\left[\frac{2k}{\sqrt{3} H_0} \left(e^{\frac{N}{2}}-e^{\frac{\tN}{2}} \right) \right]  \left(\beta_1 - \beta_2 \frac{k^2}{H_0^2}e^{\tN}\right)  \; d\tN, \label{agint} 
\eea
and we have defined
\bea
\beta_1&=& \frac{A}{2} - A^2 , \nonumber \\
\beta_2&=&\left(  \frac{1}{2}\alpha  \left(2A +1 \right) + \frac{2}{3}\left( A+2\right) \right).
\eea

Let $\omega\equiv \frac{2}{\sqrt{3} H_0}$ and then \eqref{agint} becomes
\bea
\delta_r^{(p)} &=& - \frac{\sqrt{3}H_0}{k}e^{-AN} \Phi_0 \int_0^N  e^{(A-\frac{1}{2})\tN} \sin\left[\omega k \left(e^{\frac{N}{2}}-e^{\frac{\tN}{2}} \right) \right]  \left(\beta_1 - \beta_2 \frac{k^2}{H_0^2}e^{\tN}\right)  \; d\tN, \nonumber \\
&=& \frac{\sqrt{3}H_0}{k}e^{-AN}  \cos\left(\omega k e^{\frac{N}{2}} \right) \Phi_0 \int_0^N  e^{(A-\frac{1}{2})\tN}  \sin\left(\omega k e^{\frac{\tN}{2}} \right)
 \left(\beta_1 - \beta_2 \frac{k^2}{H_0^2}e^{\tN}\right) \; d\tN \nonumber \\
&& - \frac{\sqrt{3}H_0}{k}e^{-AN}  \sin\left(\omega k e^{\frac{N}{2}} \right) \Phi_0 \int_0^N  e^{(A-\frac{1}{2})\tN}   \cos\left(\omega k e^{\frac{\tN}{2}} \right)  \left(\beta_1 - \beta_2 \frac{k^2}{H_0^2}e^{\tN}\right)  \; d\tN,
\eea

We need to solve the integrals
\bea
I_1&=& \int_0^N  e^{a\tN}  \sin\left(b e^{\frac{\tN}{2}} \right) \; d\tN \\
I_2&=& \int_0^N  e^{a\tN}  \cos\left(b e^{\frac{\tN}{2}} \right) \; d\tN \\
\eea
consider the change of variables, 
\bea
x&=&b e^{\frac{\tN}{2}}, \\
dx&=&\frac{b}{2} e^{\frac{\tN}{2}} d\tN \longrightarrow d\tN = \frac{2 dx}{x}
\eea
we then have
\bea
I_1=\int_{x(0)}^{x(N)}  \left( \frac{x}{b} \right)^{2a} \sin x \; \left(  \frac{2 dx}{x} \right)=c \int_{x(0)}^{x(N)}  x^{m} \sin x \, dx, \\
I_2=\int_{x(0)}^{x(N)}  \left( \frac{x}{b} \right)^{2a} \cos x \; \left(  \frac{2 dx}{x} \right)=c \int_{x(0)}^{x(N)}  x^{m} \cos x \, dx, 
\eea
where $c\equiv 2 b^{-2a}$ and $m\equiv 2a-1$.
We then have
\bea
\int  x^{m} \sin x \, dx &=& -\frac{i^{m+1}}{2}\left[ \Gamma\left(m+1,-ix\right) - (-1)^m  \Gamma\left(m+1,ix\right)  \right] \\
\int x^{m} \cos x \, dx &=& -\frac{i^{m+1}}{2}\left[ \Gamma\left(m+1,-ix\right) + (-1)^m  \Gamma\left(m+1,ix\right)  \right]
\eea
where we must have $m>0$.
It is also useful to note the asymptotic form as $x \rightarrow \infty$
\be
\Gamma\left(m+1,\pm ix\right) = e^{ \mp ix} x^m (\pm i)^m \left( 1 - \frac{\pm i m}{x} + {\cal O } \left(\frac{1}{x^2}\right)  \right) 
\ee
With these solutions we can now solve for the complete solution in \eqref{fullsoln}, and this can then be used to solve for the velocity perturbations.  This solution for different values of the parameters appears in Figure \ref{fig:deltarmdu}.

\bibliographystyle{utphys}

\begin{thebibliography}{10}

\bibitem{Acharya:2008bk}
B.~Acharya, P.~Kumar, K.~Bobkov, G.~Kane, J.~Shao, and S.~Watson,
  ``{Non-thermal Dark Matter and the Moduli Problem in String Frameworks},''
  \href{http://dx.doi.org/10.1088/1126-6708/2008/06/064}{{\em JHEP} {\bfseries
  0806} (2008) 064},
\href{http://arxiv.org/abs/0804.0863}{{\ttfamily arXiv:0804.0863 [hep-ph]}}.
%%CITATION = ARXIV:0804.0863;%%.

\bibitem{Acharya:2009zt}
B.~S. Acharya, G.~Kane, S.~Watson, and P.~Kumar, ``{A Non-thermal WIMP
  Miracle},'' \href{http://dx.doi.org/10.1103/PhysRevD.80.083529}{{\em
  Phys.Rev.} {\bfseries D80} (2009) 083529},
\href{http://arxiv.org/abs/0908.2430}{{\ttfamily arXiv:0908.2430
  [astro-ph.CO]}}.
%%CITATION = ARXIV:0908.2430;%%.

\bibitem{Watson:2009hw}
S.~Watson, ``{Reevaluating the Cosmological Origin of Dark Matter},'' in {\em
  Perspectives on supersymmetry II}, G.~L. Kane, ed., pp.~305--324.
\newblock World Scientific, Singapore, 2009.
\newblock
\href{http://arxiv.org/abs/0912.3003}{{\ttfamily arXiv:0912.3003 [hep-th]}}.
\newblock
%%CITATION = ARXIV:0912.3003;%%.

\bibitem{Kribs:2013lua}
G.~D. Kribs, A.~Martin, and A.~Menon, ``{Natural Supersymmetry and Implications
  for Higgs physics},''
  \href{http://dx.doi.org/10.1103/PhysRevD.88.035025}{{\em Phys.Rev.}
  {\bfseries D88} (2013) 035025},
\href{http://arxiv.org/abs/1305.1313}{{\ttfamily arXiv:1305.1313 [hep-ph]}}.
%%CITATION = ARXIV:1305.1313;%%.

\bibitem{Arvanitaki:2013yja}
A.~Arvanitaki, M.~Baryakhtar, X.~Huang, K.~Van~Tilburg, and G.~Villadoro,
  ``{The Last Vestiges of Naturalness},''
\href{http://arxiv.org/abs/1309.3568}{{\ttfamily arXiv:1309.3568 [hep-ph]}}.
%%CITATION = ARXIV:1309.3568;%%.

\bibitem{Evans:2013jna}
J.~A. Evans, Y.~Kats, D.~Shih, and M.~J. Strassler, ``{Toward Full LHC Coverage
  of Natural Supersymmetry},''
\href{http://arxiv.org/abs/1310.5758}{{\ttfamily arXiv:1310.5758 [hep-ph]}}.
%%CITATION = ARXIV:1310.5758;%%.

\bibitem{Fan:2014txa}
J.~Fan and M.~Reece, ``{A New Look at Higgs Constraints on Stops},''
\href{http://arxiv.org/abs/1401.7671}{{\ttfamily arXiv:1401.7671 [hep-ph]}}.
%%CITATION = ARXIV:1401.7671;%%.

\bibitem{Gherghetta:2014xea}
T.~Gherghetta, B.~von Harling, A.~D. Medina, and M.~A. Schmidt, ``{The price of
  being SM-like in SUSY},''
  \href{http://dx.doi.org/10.1007/JHEP04(2014)180}{{\em JHEP} {\bfseries 1404}
  (2014) 180},
\href{http://arxiv.org/abs/1401.8291}{{\ttfamily arXiv:1401.8291 [hep-ph]}}.
%%CITATION = ARXIV:1401.8291;%%.

\bibitem{Easther:2013nga}
R.~Easther, R.~Galvez, O.~Ozsoy, and S.~Watson, ``{Supersymmetry, Nonthermal
  Dark Matter and Precision Cosmology},''
\href{http://arxiv.org/abs/1307.2453}{{\ttfamily arXiv:1307.2453 [hep-ph]}}.
%%CITATION = ARXIV:1307.2453;%%.

\bibitem{Iliesiu:2013rqa}
L.~Iliesiu, D.~J.~E. Marsh, K.~Moodley, and S.~Watson, ``{Constraining SUSY
  with Heavy Scalars -- using the CMB},''
\href{http://arxiv.org/abs/1312.3636}{{\ttfamily arXiv:1312.3636
  [astro-ph.CO]}}.
%%CITATION = ARXIV:1312.3636;%%.

\bibitem{Wells:2003tf}
J.~D. Wells, ``{Implications of supersymmetry breaking with a little hierarchy
  between gauginos and scalars},''
\href{http://arxiv.org/abs/hep-ph/0306127}{{\ttfamily arXiv:hep-ph/0306127
  [hep-ph]}}.
%%CITATION = HEP-PH/0306127;%%.

\bibitem{ArkaniHamed:2004fb}
N.~Arkani-Hamed and S.~Dimopoulos, ``{Supersymmetric unification without low
  energy supersymmetry and signatures for fine-tuning at the LHC},''
  \href{http://dx.doi.org/10.1088/1126-6708/2005/06/073}{{\em JHEP} {\bfseries
  0506} (2005) 073},
\href{http://arxiv.org/abs/hep-th/0405159}{{\ttfamily arXiv:hep-th/0405159
  [hep-th]}}.
%%CITATION = HEP-TH/0405159;%%.

\bibitem{ArkaniHamed:2004yi}
N.~Arkani-Hamed, S.~Dimopoulos, G.~Giudice, and A.~Romanino, ``{Aspects of
  split supersymmetry},''
  \href{http://dx.doi.org/10.1016/j.nuclphysb.2004.12.026}{{\em Nucl.Phys.}
  {\bfseries B709} (2005) 3--46},
\href{http://arxiv.org/abs/hep-ph/0409232}{{\ttfamily arXiv:hep-ph/0409232
  [hep-ph]}}.
%%CITATION = HEP-PH/0409232;%%.

\bibitem{Giudice:2004tc}
G.~Giudice and A.~Romanino, ``{Split supersymmetry},''
  \href{http://dx.doi.org/10.1016/j.nuclphysb.2004.11.048}{{\em Nucl.Phys.}
  {\bfseries B699} (2004) 65--89},
\href{http://arxiv.org/abs/hep-ph/0406088}{{\ttfamily arXiv:hep-ph/0406088
  [hep-ph]}}.
%%CITATION = HEP-PH/0406088;%%.

\bibitem{Arvanitaki:2012ps}
A.~Arvanitaki, N.~Craig, S.~Dimopoulos, and G.~Villadoro, ``{Mini-Split},''
  \href{http://dx.doi.org/10.1007/JHEP02(2013)126}{{\em JHEP} {\bfseries 1302}
  (2013) 126},
\href{http://arxiv.org/abs/1210.0555}{{\ttfamily arXiv:1210.0555 [hep-ph]}}.
%%CITATION = ARXIV:1210.0555;%%.

\bibitem{Hall:2011jd}
L.~J. Hall and Y.~Nomura, ``{Spread Supersymmetry},''
  \href{http://dx.doi.org/10.1007/JHEP01(2012)082}{{\em JHEP} {\bfseries 1201}
  (2012) 082},
\href{http://arxiv.org/abs/1111.4519}{{\ttfamily arXiv:1111.4519 [hep-ph]}}.
%%CITATION = ARXIV:1111.4519;%%.

\bibitem{Hall:2012zp}
L.~J. Hall, Y.~Nomura, and S.~Shirai, ``{Spread Supersymmetry with Wino LSP:
  Gluino and Dark Matter Signals},''
  \href{http://dx.doi.org/10.1007/JHEP01(2013)036}{{\em JHEP} {\bfseries 1301}
  (2013) 036},
\href{http://arxiv.org/abs/1210.2395}{{\ttfamily arXiv:1210.2395 [hep-ph]}}.
%%CITATION = ARXIV:1210.2395;%%.

\bibitem{ArkaniHamed:2012gw}
N.~Arkani-Hamed, A.~Gupta, D.~E. Kaplan, N.~Weiner, and T.~Zorawski, ``{Simply
  Unnatural Supersymmetry},''
\href{http://arxiv.org/abs/1212.6971}{{\ttfamily arXiv:1212.6971 [hep-ph]}}.
%%CITATION = ARXIV:1212.6971;%%.

\bibitem{Erickcek:2011us}
A.~L. Erickcek and K.~Sigurdson, ``{Reheating Effects in the Matter Power
  Spectrum and Implications for Substructure},''
  \href{http://dx.doi.org/10.1103/PhysRevD.84.083503}{{\em Phys.Rev.}
  {\bfseries D84} (2011) 083503},
\href{http://arxiv.org/abs/1106.0536}{{\ttfamily arXiv:1106.0536
  [astro-ph.CO]}}.
%%CITATION = ARXIV:1106.0536;%%.

\bibitem{Arcadi:2011ev}
G.~Arcadi and P.~Ullio, ``{Accurate estimate of the relic density and the
  kinetic decoupling in non-thermal dark matter models},''
  \href{http://dx.doi.org/10.1103/PhysRevD.84.043520}{{\em Phys.Rev.}
  {\bfseries D84} (2011) 043520},
\href{http://arxiv.org/abs/1104.3591}{{\ttfamily arXiv:1104.3591 [hep-ph]}}.
%%CITATION = ARXIV:1104.3591;%%.

\bibitem{Craig:2014rta}
N.~Craig and D.~Green, ``{Testing Split Supersymmetry with Inflation},''
\href{http://arxiv.org/abs/1403.7193}{{\ttfamily arXiv:1403.7193 [hep-ph]}}.
%%CITATION = ARXIV:1403.7193;%%.

\bibitem{Gelmini:2006pw}
G.~B. Gelmini and P.~Gondolo, ``{Neutralino with the right cold dark matter
  abundance in (almost) any supersymmetric model},''
  \href{http://dx.doi.org/10.1103/PhysRevD.74.023510}{{\em Phys.Rev.}
  {\bfseries D74} (2006) 023510},
\href{http://arxiv.org/abs/hep-ph/0602230}{{\ttfamily arXiv:hep-ph/0602230
  [hep-ph]}}.
%%CITATION = HEP-PH/0602230;%%.

\bibitem{Fan:2013faa}
J.~Fan and M.~Reece, ``{In Wino Veritas? Indirect Searches Shed Light on
  Neutralino Dark Matter},''
  \href{http://dx.doi.org/10.1007/JHEP10(2013)124}{{\em JHEP} {\bfseries 1310}
  (2013) 124},
\href{http://arxiv.org/abs/1307.4400}{{\ttfamily arXiv:1307.4400 [hep-ph]}}.
%%CITATION = ARXIV:1307.4400;%%.

\bibitem{Cohen:2013ama}
T.~Cohen, M.~Lisanti, A.~Pierce, and T.~R. Slatyer, ``{Wino Dark Matter Under
  Siege},'' \href{http://dx.doi.org/10.1088/1475-7516/2013/10/061}{{\em JCAP}
  {\bfseries 1310} (2013) 061},
\href{http://arxiv.org/abs/1307.4082}{{\ttfamily arXiv:1307.4082}}.
%%CITATION = ARXIV:1307.4082;%%.

\bibitem{Hryczuk:2014hpa}
A.~Hryczuk, I.~Cholis, R.~Iengo, M.~Tavakoli, and P.~Ullio, ``{Indirect
  Detection Analysis: Wino Dark Matter Case Study},''
\href{http://arxiv.org/abs/1401.6212}{{\ttfamily arXiv:1401.6212
  [astro-ph.HE]}}.
%%CITATION = ARXIV:1401.6212;%%.

\bibitem{Giudice:2000ex}
G.~F. Giudice, E.~W. Kolb, and A.~Riotto, ``{Largest temperature of the
  radiation era and its cosmological implications},''
  \href{http://dx.doi.org/10.1103/PhysRevD.64.023508}{{\em Phys.Rev.}
  {\bfseries D64} (2001) 023508},
\href{http://arxiv.org/abs/hep-ph/0005123}{{\ttfamily arXiv:hep-ph/0005123
  [hep-ph]}}.
%%CITATION = HEP-PH/0005123;%%.

\bibitem{Cheung:2010gj}
C.~Cheung, G.~Elor, L.~J. Hall, and P.~Kumar, ``{Origins of Hidden Sector Dark
  Matter I: Cosmology},'' \href{http://dx.doi.org/10.1007/JHEP03(2011)042}{{\em
  JHEP} {\bfseries 1103} (2011) 042},
\href{http://arxiv.org/abs/1010.0022}{{\ttfamily arXiv:1010.0022 [hep-ph]}}.
%%CITATION = ARXIV:1010.0022;%%.

\bibitem{Lemoine:2009is}
M.~Lemoine, J.~Martin, and J.~Yokoyama, ``{Constraints on moduli cosmology from
  the production of dark matter and baryon isocurvature fluctuations},''
  \href{http://dx.doi.org/10.1103/PhysRevD.80.123514}{{\em Phys.Rev.}
  {\bfseries D80} (2009) 123514},
\href{http://arxiv.org/abs/0904.0126}{{\ttfamily arXiv:0904.0126
  [astro-ph.CO]}}.
%%CITATION = ARXIV:0904.0126;%%.

\bibitem{Brandenberger:1992dw}
R.~H. Brandenberger, H.~Feldman, V.~F. Mukhanov, and T.~Prokopec,
``{Gauge invariant cosmological perturbations: Theory and applications},''.
%%CITATION = BROWN-HET-860 ETC.;%%.

\bibitem{Boehm:2000gq}
C.~Boehm, P.~Fayet, and R.~Schaeffer, ``{Constraining dark matter candidates
  from structure formation},''
  \href{http://dx.doi.org/10.1016/S0370-2693(01)01060-7}{{\em Phys.Lett.}
  {\bfseries B518} (2001) 8--14},
\href{http://arxiv.org/abs/astro-ph/0012504}{{\ttfamily arXiv:astro-ph/0012504
  [astro-ph]}}.
%%CITATION = ASTRO-PH/0012504;%%.

\bibitem{Chen:2001jz}
X.-l. Chen, M.~Kamionkowski, and X.-m. Zhang, ``{Kinetic decoupling of
  neutralino dark matter},''
  \href{http://dx.doi.org/10.1103/PhysRevD.64.021302}{{\em Phys.Rev.}
  {\bfseries D64} (2001) 021302},
\href{http://arxiv.org/abs/astro-ph/0103452}{{\ttfamily arXiv:astro-ph/0103452
  [astro-ph]}}.
%%CITATION = ASTRO-PH/0103452;%%.

\bibitem{Hofmann:2001bi}
S.~Hofmann, D.~J. Schwarz, and H.~Stoecker, ``{Damping scales of neutralino
  cold dark matter},'' \href{http://dx.doi.org/10.1103/PhysRevD.64.083507}{{\em
  Phys.Rev.} {\bfseries D64} (2001) 083507},
\href{http://arxiv.org/abs/astro-ph/0104173}{{\ttfamily arXiv:astro-ph/0104173
  [astro-ph]}}.
%%CITATION = ASTRO-PH/0104173;%%.

\bibitem{Bertschinger:2006nq}
E.~Bertschinger, ``{The Effects of Cold Dark Matter Decoupling and Pair
  Annihilation on Cosmological Perturbations},''
  \href{http://dx.doi.org/10.1103/PhysRevD.74.063509}{{\em Phys.Rev.}
  {\bfseries D74} (2006) 063509},
\href{http://arxiv.org/abs/astro-ph/0607319}{{\ttfamily arXiv:astro-ph/0607319
  [astro-ph]}}.
%%CITATION = ASTRO-PH/0607319;%%.

\bibitem{Green:2003un}
A.~M. Green, S.~Hofmann, and D.~J. Schwarz, ``{The power spectrum of SUSY - CDM
  on sub-galactic scales},''
  \href{http://dx.doi.org/10.1111/j.1365-2966.2004.08232.x}{{\em
  Mon.Not.Roy.Astron.Soc.} {\bfseries 353} (2004) L23},
\href{http://arxiv.org/abs/astro-ph/0309621}{{\ttfamily arXiv:astro-ph/0309621
  [astro-ph]}}.
%%CITATION = ASTRO-PH/0309621;%%.

\bibitem{Green:2005fa}
A.~M. Green, S.~Hofmann, and D.~J. Schwarz, ``{The First wimpy halos},''
  \href{http://dx.doi.org/10.1088/1475-7516/2005/08/003}{{\em JCAP} {\bfseries
  0508} (2005) 003},
\href{http://arxiv.org/abs/astro-ph/0503387}{{\ttfamily arXiv:astro-ph/0503387
  [astro-ph]}}.
%%CITATION = ASTRO-PH/0503387;%%.

\bibitem{Boyarsky:2008xj}
A.~Boyarsky, J.~Lesgourgues, O.~Ruchayskiy, and M.~Viel, ``{Lyman-alpha
  constraints on warm and on warm-plus-cold dark matter models},''
  \href{http://dx.doi.org/10.1088/1475-7516/2009/05/012}{{\em JCAP} {\bfseries
  0905} (2009) 012},
\href{http://arxiv.org/abs/0812.0010}{{\ttfamily arXiv:0812.0010 [astro-ph]}}.
%%CITATION = ARXIV:0812.0010;%%.

\bibitem{Loeb:2005pm}
A.~Loeb and M.~Zaldarriaga, ``{The Small-scale power spectrum of cold dark
  matter},'' \href{http://dx.doi.org/10.1103/PhysRevD.71.103520}{{\em
  Phys.Rev.} {\bfseries D71} (2005) 103520},
\href{http://arxiv.org/abs/astro-ph/0504112}{{\ttfamily arXiv:astro-ph/0504112
  [astro-ph]}}.
%%CITATION = ASTRO-PH/0504112;%%.

\bibitem{Lin:2000qq}
W.~Lin, D.~Huang, X.~Zhang, and R.~H. Brandenberger, ``{Nonthermal production
  of WIMPs and the subgalactic structure of the universe},''
  \href{http://dx.doi.org/10.1103/PhysRevLett.86.954}{{\em Phys.Rev.Lett.}
  {\bfseries 86} (2001) 954},
\href{http://arxiv.org/abs/astro-ph/0009003}{{\ttfamily arXiv:astro-ph/0009003
  [astro-ph]}}.
%%CITATION = ASTRO-PH/0009003;%%.

\bibitem{Moroi:1999zb}
T.~Moroi and L.~Randall, ``{Wino cold dark matter from anomaly mediated SUSY
  breaking},'' \href{http://dx.doi.org/10.1016/S0550-3213(99)00748-8}{{\em
  Nucl.Phys.} {\bfseries B570} (2000) 455--472},
\href{http://arxiv.org/abs/hep-ph/9906527}{{\ttfamily arXiv:hep-ph/9906527
  [hep-ph]}}.
%%CITATION = HEP-PH/9906527;%%.

\bibitem{Bhattacherjee:2014dya}
B.~Bhattacherjee, M.~Ibe, K.~Ichikawa, S.~Matsumoto, and K.~Nishiyama, ``{Wino
  Dark Matter and Future dSph Observations},''
\href{http://arxiv.org/abs/1405.4914}{{\ttfamily arXiv:1405.4914 [hep-ph]}}.
%%CITATION = ARXIV:1405.4914;%%.

\bibitem{Allahverdi:2013noa}
R.~Allahverdi, M.~Cicoli, B.~Dutta, and K.~Sinha, ``{Non-thermal Dark Matter in
  String Compactifications},'' {\em Phys.Rev.} {\bfseries D88} (2013) 095015,
\href{http://arxiv.org/abs/1307.5086}{{\ttfamily arXiv:1307.5086 [hep-ph]}}.
%%CITATION = ARXIV:1307.5086;%%.

\bibitem{Ibe:2012sx}
M.~Ibe, S.~Matsumoto, and R.~Sato, ``{Mass Splitting between Charged and
  Neutral Winos at Two-Loop Level},''
  \href{http://dx.doi.org/10.1016/j.physletb.2013.03.015}{{\em Phys.Lett.}
  {\bfseries B721} (2013) 252--260},
\href{http://arxiv.org/abs/1212.5989}{{\ttfamily arXiv:1212.5989 [hep-ph]}}.
%%CITATION = ARXIV:1212.5989;%%.

\bibitem{ArkaniHamed:2006mb}
N.~Arkani-Hamed, A.~Delgado, and G.~Giudice, ``{The Well-tempered
  neutralino},'' \href{http://dx.doi.org/10.1016/j.nuclphysb.2006.02.010}{{\em
  Nucl.Phys.} {\bfseries B741} (2006) 108--130},
\href{http://arxiv.org/abs/hep-ph/0601041}{{\ttfamily arXiv:hep-ph/0601041
  [hep-ph]}}.
%%CITATION = HEP-PH/0601041;%%.

\bibitem{Hisano:2000dz}
J.~Hisano, K.~Kohri, and M.~M. Nojiri, ``{Neutralino warm dark matter},''
  \href{http://dx.doi.org/10.1016/S0370-2693(01)00395-1}{{\em Phys.Lett.}
  {\bfseries B505} (2001) 169--176},
\href{http://arxiv.org/abs/hep-ph/0011216}{{\ttfamily arXiv:hep-ph/0011216
  [hep-ph]}}.
%%CITATION = HEP-PH/0011216;%%.

\bibitem{Bardeen:1983qw}
J.~M. Bardeen, P.~J. Steinhardt, and M.~S. Turner, ``{Spontaneous Creation of
  Almost Scale - Free Density Perturbations in an Inflationary Universe},''
  \href{http://dx.doi.org/10.1103/PhysRevD.28.679}{{\em Phys.Rev.} {\bfseries
  D28} (1983) 679}.

\bibitem{Linde:2005ht}
A.~D. Linde, ``{Particle physics and inflationary cosmology},'' {\em
  Contemp.Concepts Phys.} {\bfseries 5} (1990) 1--362,
\href{http://arxiv.org/abs/hep-th/0503203}{{\ttfamily arXiv:hep-th/0503203
  [hep-th]}}.
%%CITATION = HEP-TH/0503203;%%.

\bibitem{Malik:2004tf}
K.~A. Malik and D.~Wands, ``{Adiabatic and entropy perturbations with
  interacting fluids and fields},''
  \href{http://dx.doi.org/10.1088/1475-7516/2005/02/007}{{\em JCAP} {\bfseries
  0502} (2005) 007},
\href{http://arxiv.org/abs/astro-ph/0411703}{{\ttfamily arXiv:astro-ph/0411703
  [astro-ph]}}.
%%CITATION = ASTRO-PH/0411703;%%.

\end{thebibliography}

\providecommand{\href}[2]{#2}\begingroup\raggedright\endgroup

\end{document}